\documentclass[9pt,twocolumn,twoside]{osajnl}

\journal{ol} 

\setboolean{shortarticle}{true}

\title{Geometrical and physical optics analysis for mm-wavelength refractor telescopes designed to map the cosmic microwave background}
\author[1]{$^*$Jon E. Gudmundsson}
\affil[1]{The Oskar Klein Centre,
Department of Physics, Stockholm University, SE-106 91 Stockholm, Sweden}

\affil[*]{Contact author via: jon@fysik.su.se}

\begin{abstract}
We present a compact two-lens HDPE $f$/1.6 refractor design that is capable of supporting a 28-deg diffraction-limited field of view at 1-mm wavelengths and contrast it to a similar two-lens refractor using silicon lenses. We compare the optical properties of these two systems as predicted by both geometrical and physical optics. The presented analysis suggests that by relaxing telecentricity requirements, a plastic two-lens refractor system can perform comparably to a similar silicon system across a wide field of view and wavelengths up to 1 mm. We show that for both telescope designs, cold stop spillover changes significantly across the field of view in a way that is somewhat inconsistent with Gaussian beam formalism and simple $f$-number scaling. We present results that highlight beam ellipticity dependence on both pixel location and pixel aperture size---an effect that is challenging to reproduce in standard geometrical optics. We show that a silicon refractor design suffers from larger cross-polarization compared to the HDPE design. Our results address the limitations of solely relying on geometrical optics to assess relative performance of two optical systems. We discuss implications for future refractor designs.
\end{abstract}

\setboolean{displaycopyright}{false}
\newcommand{\mrm}[1]{\mathrm{#1}} 
\usepackage{xcolor}

\begin{document}
\maketitle

\noindent Observations at mm-wavelengths are used to constrain a range of astrophysical phenomena, including: various cosmological signals embedded in the cosmic microwave background (CMB) \cite{Planck2018_parameters}, emission from dust and synchrotron radiation in the Milky Way \cite{Planck2018_foregrounds}, protoplanetary disks in our stellar neighborhood \cite{ALMA2015}, and even supermassive black holes in nearby galaxies \cite{EHT2019}. Many of those measurements are limited by the optical noise from either atmospheric or instrument thermal emission. This is particularly true for CMB experiments that require long observation times to extract a minute polarized signal from largely noise-dominated detector timestreams. In an effort to enhance overall sensitivity, the CMB community is developing telescope designs that facilitate dramatic increase in optical throughput. The majority of these new designs rely on refractive optical elements \cite{Niemack2016, Galitzki2018, Hui2018} to some extent. This push for high throughput refractive designs is bringing about modeling challenges that need to be addressed in order to accurately forecast detector sensitivities and systematic effects. The purpose of this paper is to present results from a fast optical analysis framework that combines geometrical and physical optics to compare two compact refractive telescopes designed to map the sky at mm-wavelengths.

Instruments designed for observations of the CMB use refractors either as standalone telescopes \cite{Takahashi2010, Hui2018, Galitzki2018} or as reimaging optics coupling to reflectors \cite{Ruhl2004, Fowler2007, Niemack2016, Parshley2018}. Typical lens materials incorporated in these refractors include plastics (index of refraction $n \approx 1.5$), such as high-density polyethylene (HDPE) or polypropylene (PP) \cite{Runyan2010, DAlessandro2018}, as well as higher index of refraction materials alumina ($n \approx 3.1$) \cite{Ahmed2014} and silicon ($n \approx 3.4$) \cite{Thornton2016, Galitzki2018}. In this paper, we compare two 30-cm diameter, $f$/1.6 on-axis refractor designs; one using HDPE lenses and another using silicon lenses. The two designs are optimized using the same performance metrics and made to have very similar effective $f$-numbers across the field of view (FOV). However, the high index of refraction of silicon would be expected to facilitate greater optical performance overall. We choose to simulate relatively compact 30-cm diameter telescopes, but note that many of the results presented in this paper can be used to inform larger, higher throughput (étendue), telescope designs. The two designs presented in this paper were developed with ballooning and space missions in mind. Consequently, we tried to minimize mass and volume by keeping lenses as thin as possible and by constraining lens diameters so that they did not significantly exceed the aperture diameter. For the same reason, we tried to maximize field of view per unit telescope volume by using fast (low $f$-number) optics. Similar optical designs could be used for balloon-borne or satellite experiments meant to study CMB polarization on the largest angular scales.

By relaxing a telecentricity requirement, we show that at 150 GHz, a 30-cm diameter two-lens HDPE design can support a diffraction-limited optical throughput of 132~srad~cm$^2$. This is a factor of 3 improvement over deployed small aperture ($\lesssim 0.5$~m) CMB telescopes that use plastic lenses \cite{Hanany2013}. The physical optics analysis presented in this paper indicates that reduced telecentricity does not significantly diminish far field beam fidelity. Replacing the plastic lenses with high index of refraction materials such as silicon or alumina, can lead to even greater optical throughput for simple two-lens telescopes constrained to fit within a fixed volume.

This paper is structured as follows. In Section \ref{sec:throughput} we discuss telescope throughput to motivate the paper. Section \ref{sec:design} describes the basic design philosophy for the two refractor telescopes and discusses the optimization process. Section \ref{sec:go} presents basic geometrical optics properties of the two telescopes and simple tolerancing analysis. Section \ref{sec:po} shows physical optics simulations results covering properties such as beam solid angle, ellipticity, spillover, and cross-polarization. Section \ref{sec:gopocomp} discusses the prospects for comparing the geometrical and physical optical results. Finally, we follow this with a general discussion and conclusions in Sections \ref{sec:discussion} and \ref{sec:conclusions}, respectively.

\section{Refractor throughput}
\label{sec:throughput}

\begin{figure}
\begin{center}
\includegraphics[width=9cm]{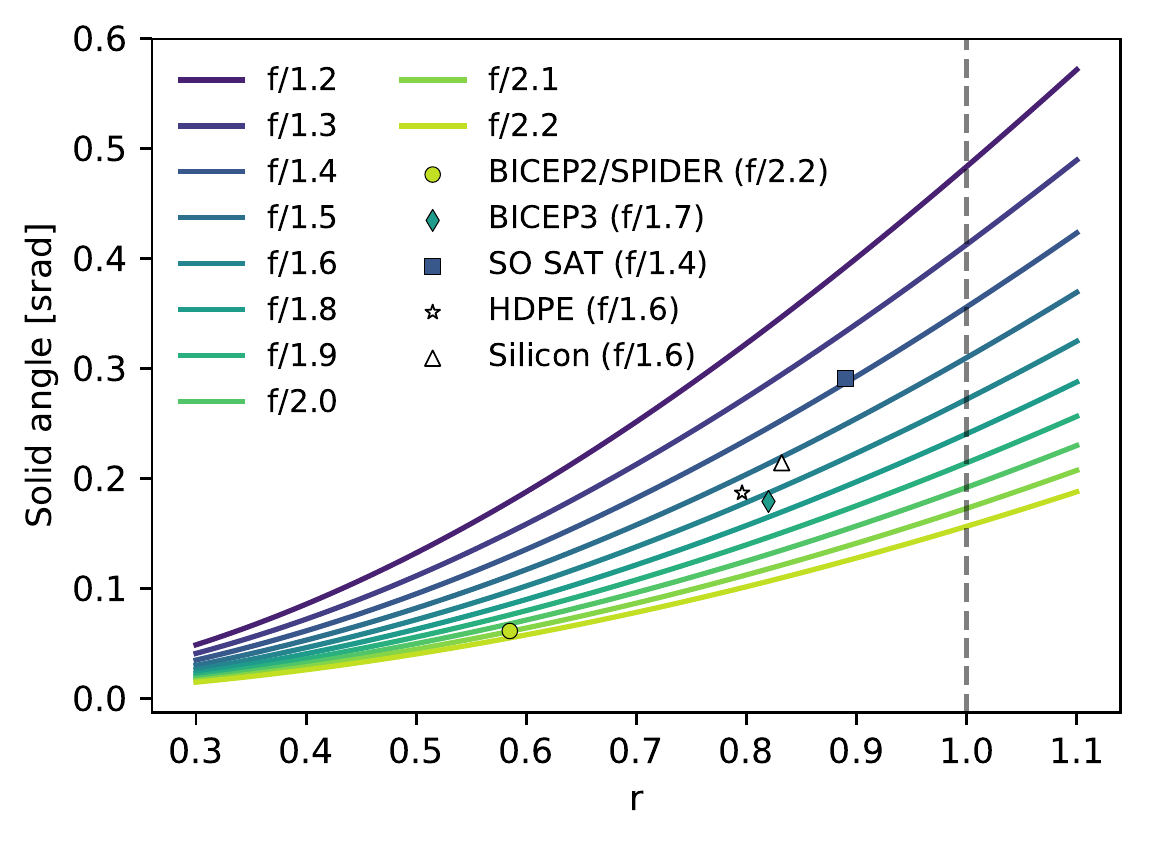}
\caption{Diffraction-limited telescope solid angle as a function of the ratio between focal plane diameter and telescope aperture, $r$ (see Equation \ref{eq:omega}). The two telescopes that are discussed in this paper, the HDPE and silicon designs, are shown with star- and triangle-shaped markers, respectively. Three existing refractor telescope designs are also shown \cite{Aikin2010, Runyan2010, Wu2015, Galitzki2018}. For those three telescope designs, we only show the solid angle that is covered by the experiment, not the diffraction-limited solid angle (although those can be very similar).} 
\label{fig:sa}
\end{center}
\end{figure}

Design challenges for contemporary CMB telescopes include a tradeoff between minimizing systematics and maximizing optical throughput; a challenge partly caused by the fact that optical systematics are reduced in slow systems with significant absorptive baffling that in turn increases optical loading on detectors~\cite{Griffin2002}. To achieve their science goals, current and future CMB experiments are pushing the limits of available mapping speed or equivalently,  the instantaneous sensitivity of the experiment integrated over a fixed amount of time. Mapping speed can be increased by improving the optical throughput \cite{Niemack2016} or by simply building multiple copies of the same telescope. 

\begin{figure}
\begin{center}
\includegraphics[width=8cm]{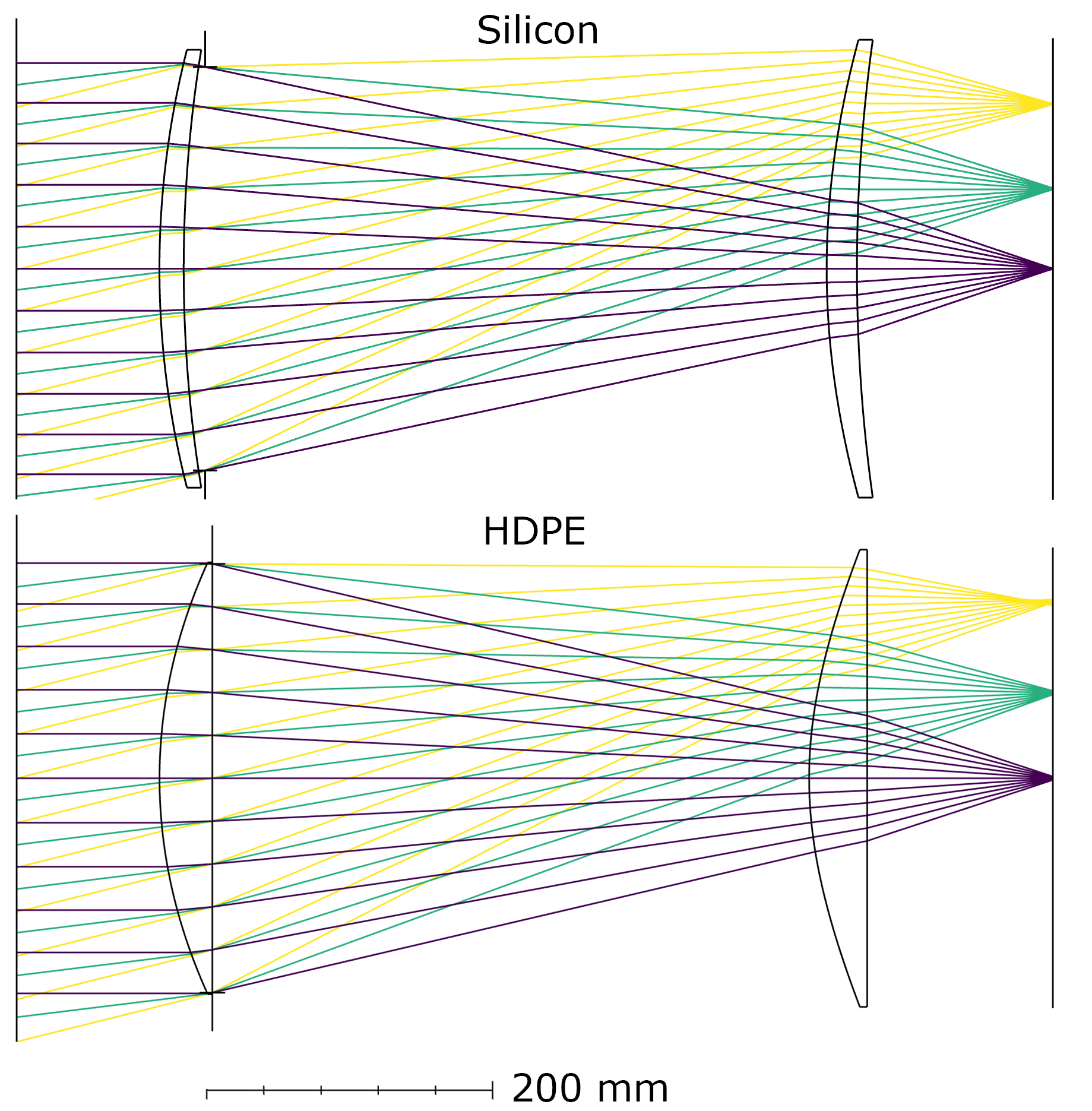}
\caption{Time-forward ray trace diagram of the proposed silicon (top) and HDPE (bottom) design. The three bundles of rays, blue, green and yellow, correspond to the center, middle, and edge of the half field of view (14~deg), respectively. The distance from the front of the primary lens to the focal plane is about 625~mm. Table \ref{tab:opt_prop} lists other dimensional properties of the two designs.}
\label{fig:ray_trace}
\end{center}
\end{figure}

In an ideal situation where all radiation modes entering the system are absorbed in detectors, the diffraction-limited optical throughput, $A\Omega$, can be related to the number of modes through

\begin{equation}
N \lambda ^2 = A\Omega =  2\pi A \left(1-\cos \theta _{\mrm{FOV}} \right),
\end{equation}
where $N$ is the total number of modes, $A$ is the entrance pupil (aperture) area, $\Omega$ is the solid angle subtended by the telescope, and $\theta_{\mrm{FOV}}$ is the half field of view \cite{Kraus1986}. For a focal plane with diameter $d$, we can write 

\begin{equation}
d = r D, 
\end{equation}
where $r$ is a scaling factor that corresponds to the relative dimensions of the focal plane diameter, $d$, and the aperture diameter, $D$.

Using the well known relation connecting the field of view, $\theta _{\mrm{FOV}}$, focal length, $L$, and the focal plane diameter, $d$,
\begin{equation}
\theta _{\mrm{FOV}} = \arctan \left( \frac{d}{2L} \right),
\end{equation}
one can write
\begin{equation}
\Omega = 2\pi \left[1-\cos \left( \arctan \left( r/2F \right) \right) \right], 
\label{eq:omega}
\end{equation}
assuming $F = L/D$, the telescope $f$-number, remains constant across the field of view. This expression is instructive since it relates the telescope FOV with both $f$-number and focal plane radius. Unfortunately, in most cases the telescope $f$-number grows with field location and therefore Equation \ref{eq:omega} can only be taken as approximate. Nevertheless, Figure \ref{fig:sa} shows the telescope solid angle for a family of $f$-number curves as a function of $r$. We compare these curves against the utilized solid angle for the BICEP2/SPIDER \cite{Aikin2010, Runyan2010}, BICEP3 \cite{Wu2015}, and Simons Observatory Small Aperture Telescope \cite{Galitzki2018}. The BICEP2/SPIDER telescope was first deployed in 2010, the BICEP3 telescope in 2015, and the Simons Observatory SAT telescope is expected to deploy in 2021. Small aperture telescopes for CMB observations are clearly pushing for increasing telescope field of view.

It is important to note that Figure \ref{fig:sa} does not express the optical throughput, $A\Omega$, which differs significantly between experiments. Since diffraction effects scale non-trivially with overall dimensions, blind scaling of existing telescope designs is often unrealistic. The BICEP3 telescope (52 cm aperture) and Simons Observatory small aperture telescope (40 cm aperture) report a 150-GHz diffraction-limited optical throughput of 380 and 400~cm$^2$~srad, respectively \cite{BICEP2K_2015_IV, Galitzki2018}. This is significantly larger than that of the BICEP2/SPIDER telescope (40 ~cm$^2$~srad) and the two telescope models that are described in this paper (roughly 140~cm$^2$~srad). However, given that a 30-cm plastic-lens refractor can support 140~cm$^2$~srad at mm-wavelengths, it is reasonable to assume that an even larger throughput can be supported by slightly increasing the aperture size while making modest changes to the lens configuration.




\section{Optical design}
\label{sec:design}

\begin{table}
\caption{Optical properties of the two designs considered in this paper. Note that $c$ and $k$ represent the inverse radius of curvature and the conic constant, respectively. The silicon lenses are assumed to have an index of refraction of $n_\mrm{si} = 3.42$ and the physical separation between primary and secondary lens is 450 mm. The axial thickness is 17.0 and 21.2~mm for the primary and secondary, respectively. The focal plane is located 137 mm behind the secondary lens. The HDPE lenses are assumed to have an index of refraction of $n_\mrm{HDPE} = 1.52$ and the axial separation between primary and secondary lens is 400 mm. The axial thickness is 36.9 and 40.0~mm for the primary and secondary, respectively. The focal plane is located 130 mm behind the secondary lens.\label{tab:opt_prop}}

\begin{tabular}{lllll}
\hline
Lens & surface & $c$ [m$^{-1}$] & $k$ \\
\hline
\textbf{Silicon Primary} & Front (sky) & 1.446 & 0.141  \\
 & Back & 0.933 & 1.369 \\
\textbf{Silicon Secondary} & Front & 1.635 & -0.052  \\
 & Back & 0.834 & 14.841  \\
\textbf{HDPE Primary} & Front & 2.908 & -0.670  \\
\textbf{HDPE Secondary} & Front & 3.340 & -4.439 \\
\hline
\end{tabular}
\end{table}

Ray tracing diagrams of the two proposed designs are shown in Figure \ref{fig:ray_trace} while Table \ref{tab:opt_prop} presents the primary design parameters of the two systems. The two designs are chosen to be quite similar. Both employ two aspheric 30-cm diameter lenses with a cold stop that is placed just inside of the primary lens. We note that Figure \ref{fig:ray_trace} omits any internal baffling, thermal filters, or apertures formed by intermediate cryogenic stages. Deployment of these kinds of systems on a CMB experiment would almost certainly mean that both lenses were cooled down to roughly 4--10~K in a configuration that incorporated both a cooled optics sleeve and forebaffles. Zemax files of the two designs can be accessed from links in the following footnote.\footnote{\href{https://github.com/jegudmunds/refractor\_designs}{https://github.com/jegudmunds/refractor\_designs}}

\begin{figure}
\begin{center}
\includegraphics[width=9cm]{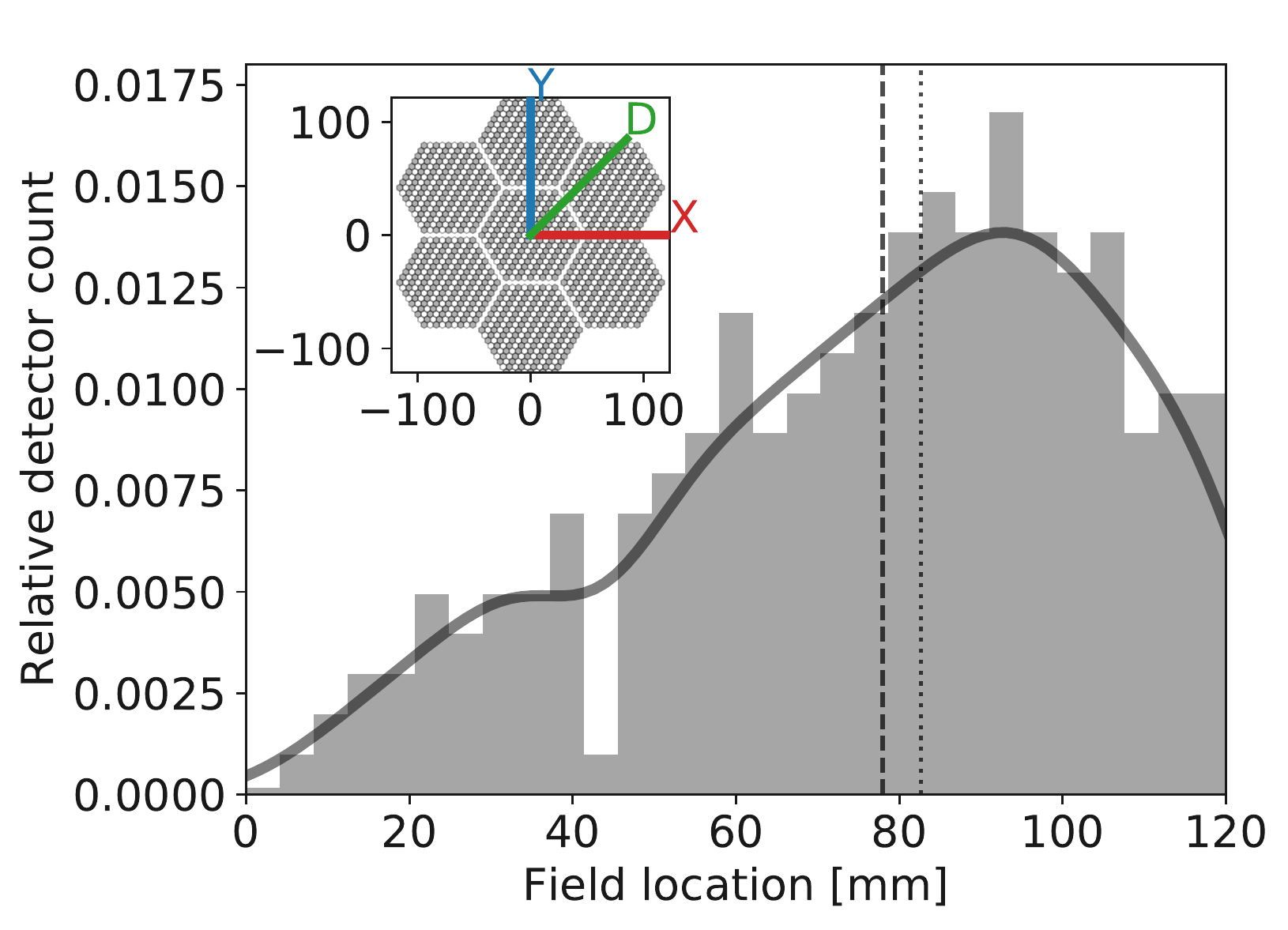}
\caption{Approximate pixel distribution for a 240-mm diameter focal plane populated with seven hexagonal tiles. Each of the detector tiles has a 84-mm diameter and can support 217 physical pixels. The mean and the median of the distribution are shown with the dashed and dotted lines, respectively. The focal plane layout is shown in the inset figure. The red, green, and blue curves in the inset figure are used to outline three principal axes used in subsequent analysis, labeled X, Y, and D.}
\label{fig:distribution}
\end{center}
\end{figure}

\subsection{Optimization}

The lens optimization began with aspheric two-lens telescopes with dimensional constraints such that the system working $f$-number took a value of approximately $f$/1.6. The thicknesses of the two lenses as well as the spacing between lenses and the focal plane were allowed to vary while the aperture stop was fixed to coincide with the inside of the primary lens. With the radius of curvature and conic constants of four surfaces set as free parameters, the overall optimization process had a total of 11 free parameters (since the $f$-number was constrained). The optimization process began with a reduced number of free parameters to identify classes of viable designs before exploring the full range of free parameters. The thickness of any lens element was constrained to stay within about 40~mm. 

The design merit function was configured so that we would optimize Strehl ratios across the 28-deg field of view while maximizing telecentricity and keeping $f$-number variations at a minimum. In the case of the HDPE design, the optimization process seemed to converge on relatively flat back surface for both the primary and secondary lenses. For the purpose of simplifying the optimization of the lens, we chose to constrain the back surfaces to be planar (see Figure \ref{fig:ray_trace}). This obviously reduced the number of free parameters in the optimization process by four. 

The optimization process incorporated field weighting that was representative of the expected pixel distribution across the focal plane (see Figure \ref{fig:distribution}). This approach helps maximize optical performance in the field region that has the greatest pixel density. We note that a focal plane constructed out of seven 84-mm hexagonal detector tiles with 6-mm pixel horn apertures should be able to support approximately 1500 physical detectors. By employing dichroic pixels with A/B polarization pairs the focal plane could therefore support about 6000 independent channels. Figure \ref{fig:distribution} shows the normalized pixel distribution as a function of focal plane location for this scenario. 

Our loosely defined requirements and performance metric leave significant room for improvement of the two designs. In particular, we expect that a two-lens silicon design can be made to support an even wider diffraction-limited field field of view. Also, the inclusion of higher order aspheric terms would have improved the designs, but we chose to forego such additions to save time and reduce complexity.  

\subsection{Comparison of HDPE and silicon designs}
With an index of refraction of approximately 1.5, easily manufactured plastics such as HDPE and PP make for attractive lens materials. Plastic lenses have arguably slightly higher technological readiness level (TRL) compared to alumina and silicon as they have been deployed on a number of ground \cite{Iuliano2018} and balloon-borne \cite{Misawa2014, Rahlin2014} missions. Plastic lenses were also used as part of the horn assembly on the Planck satellite \cite{Planck2010_optical_architecture}. 

Plastic lenses are generally inexpensive and easy to manufacture compared to silicon. On the other hand, their relatively poor thermal conductivity and high opacity can lead to increased detector loading. Because of the relatively low index of refraction, anti-reflection coatings for plastics can theoretically support quite broad frequency ranges compared to silicon. The low mm-wavelength index of refraction also suggests that it is difficult to construct simple optical designs that support wide FOV's while remaining  telecentric. 

Silicon is seeing increasing use in mm-wavelength experiments largely due to technological advancements in anti-reflection coating techniques \cite{Niemack2010, Galitzki2018}. The high index of refraction ($n\approx3.4$), low loss, and good thermal conductivity make it a particularly appealing lens material for cryogenic optics. Because of its high index of refraction, however, anti-reflection coating becomes crucial. This has been achieved by cutting a succession of grooves in the surface of the lenses \cite{Datta2013}. Both silicon and alumina also are expected to provide improved dimensional stability over HDPE since their coefficient of thermal contraction is considerably lower than that of plastic \cite{S4_2017_technology_book}. 

Fig.~\ref{fig:ray_trace} shows a ray tracing diagram of the proposed two-lens refractor. The designs have an effective $f$-number spanning 1.6--1.8. The corresponding Strehl ratios at 90, 150, 280, and 350 GHz for the two designs are shown in Fig.~\ref{fig:strehl_ratio}.

\section{Geometrical properties}
\label{sec:go}

We present some basic results of geometrical optics analysis for the two designs as calculated using Zemax OpticStudio \cite{Zemax}.\footnote{These calculations were performed with Zemax 18.1.}

\begin{figure}
\begin{center}
\includegraphics[width=9cm]{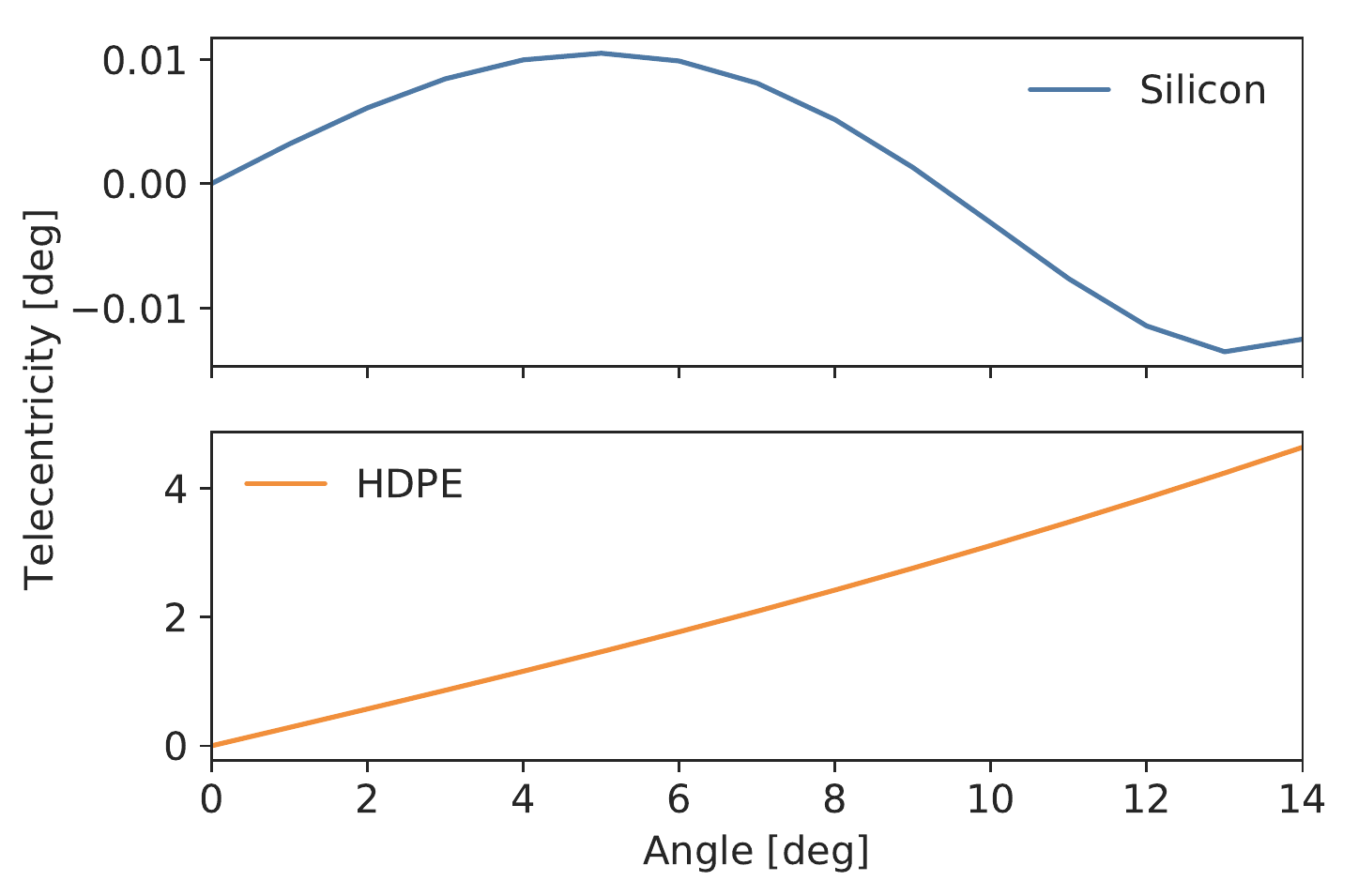}
\caption{Telecentricity across the field for the two designs. The silicon design is telecentric across the entire field of view while the HDPE design is not.}
\label{fig:telecentricity}
\end{center}
\end{figure}

\subsection{Telecentricity}
The telecentricity angle is the angle of incidence that the chief ray makes with the focal plane. This property quantifies the telescope aperture stop illumination symmetry which in turn impacts the symmetry of the telescope beam response on the sky. The two designs differ significantly in terms of telecentricity. The silicon design is telecentric to within 0.1~deg over the whole field of view while the incident angle of the chief ray grows to about 4.4~deg at the edge of the field for the HDPE design. Figure \ref{fig:telecentricity} shows the telecentricity of the two designs across the half FOV. One of the main goals of this paper is to see if this relaxing of the telecentricity requirement leads to significantly reduced far field beam performance (see Section \ref{sec:po}). 

\subsection{Strehl ratio}

The Strehl ratio, represents the squared and time-averaged phase amplitude error averaged across a focal plane image:
\begin{equation}
S_\mrm{r} = | \langle e^{i 2\pi\delta / \lambda} \rangle |^2.
\label{eq:sr}
\end{equation}
It is often stated that a "diffraction-limited" image has a Strehl ratio exceeding 0.8 (the Maréchal criterion) which roughly corresponds to an average wavefront error of about 1/14 wavelengths. Equation \ref{eq:sr} can of course be related to telescope phase and aperture efficiencies, which are figures of merit that are commonly used in the design of radio instruments \cite{Ruze1966, Olmi2007}. The Strehl ratio is also one of the first property used to compare designs of telescopes operating at mm-wavelengths \cite{S4_2017_technology_book}.

\begin{figure}
\begin{center}
\includegraphics[width=9cm]{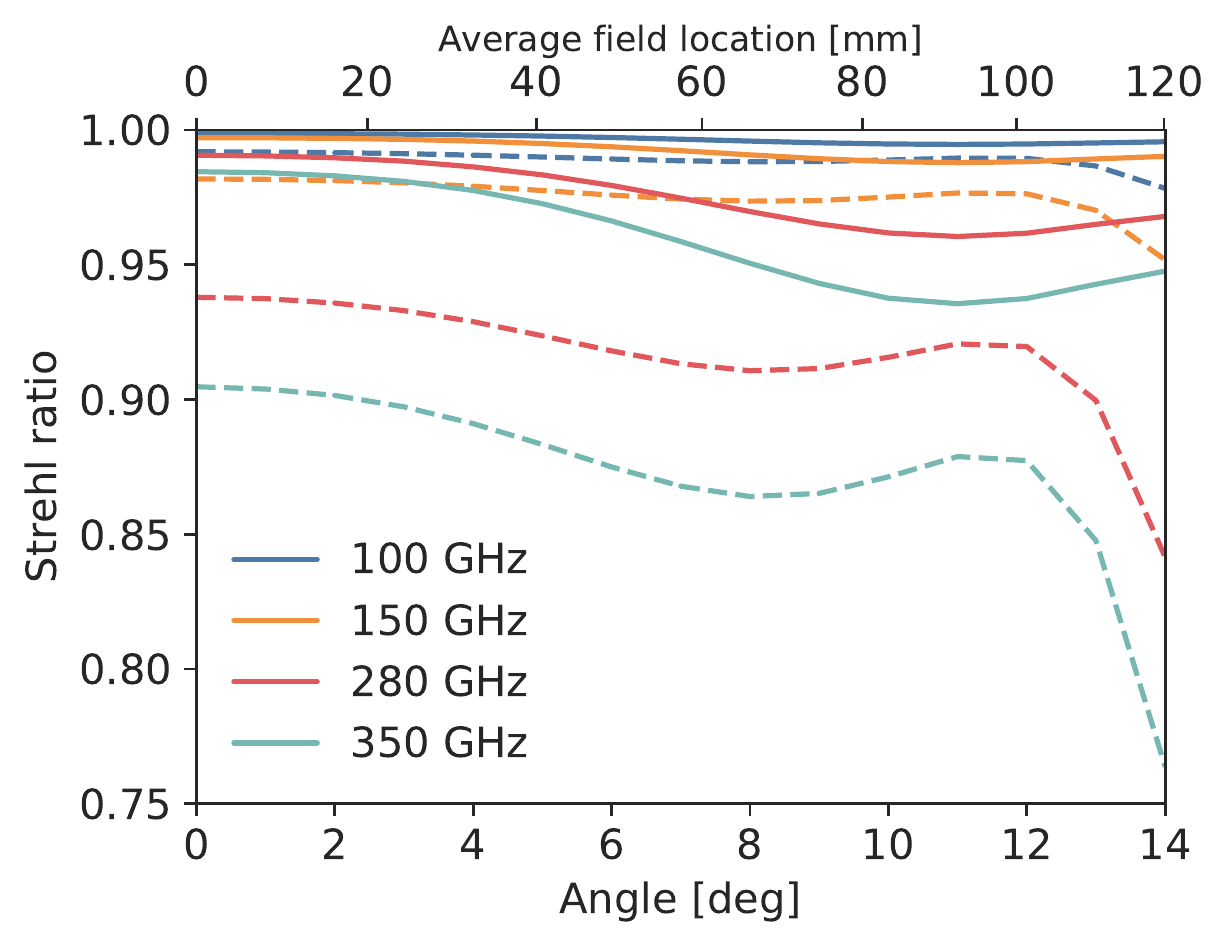}
\caption{Strehl ratio across the field for the two proposed designs. From top to bottom, blue, orange, red, and teal curves correspond to 100, 150, 280, and 350~GHz, respectively. Solid and dashed lines correspond to the silicon and HDPE designs, respectively. Clearly, the HDPE design is not able to support high frequency channels much beyond 14~degrees, but the 350-GHz Strehl ratio for the silicon design is still well above 0.8 at 14 degrees.}
\label{fig:strehl_ratio}
\end{center}
\end{figure}

Figure \ref{fig:strehl_ratio} shows the time-forward Strehl ratio at 90, 150, 280, and 350~GHz for the two designs. The two designs have diffraction-limited field of view exceeding 28 degrees for frequencies up to 350~GHz (0.857~mm), but we note that the Strehl ratio starts to decline rapidly at around 12 degree field angles for the HDPE design. The two designs have quite different Strehl ratios across the FOV and we are interested in seeing if and how the lower Strehl ratio of the HDPE design leads to significantly lower predicted far field beam performance (see Section \ref{sec:po}).\footnote{Since the HDPE design is not perfectly telecentric, we note that it can be useful to look at the time-reverse Strehl ratio, which we define as the average phase error integrated across the aperture plane when launching a bundle of rays from the focal plane. We find that the time-reverse Strehl ratios are quite comparable to the time-forward Strehl ratios.}

\subsection{Effective $f$-number}
\begin{figure}[t]
\begin{center}
\includegraphics[width=9cm]{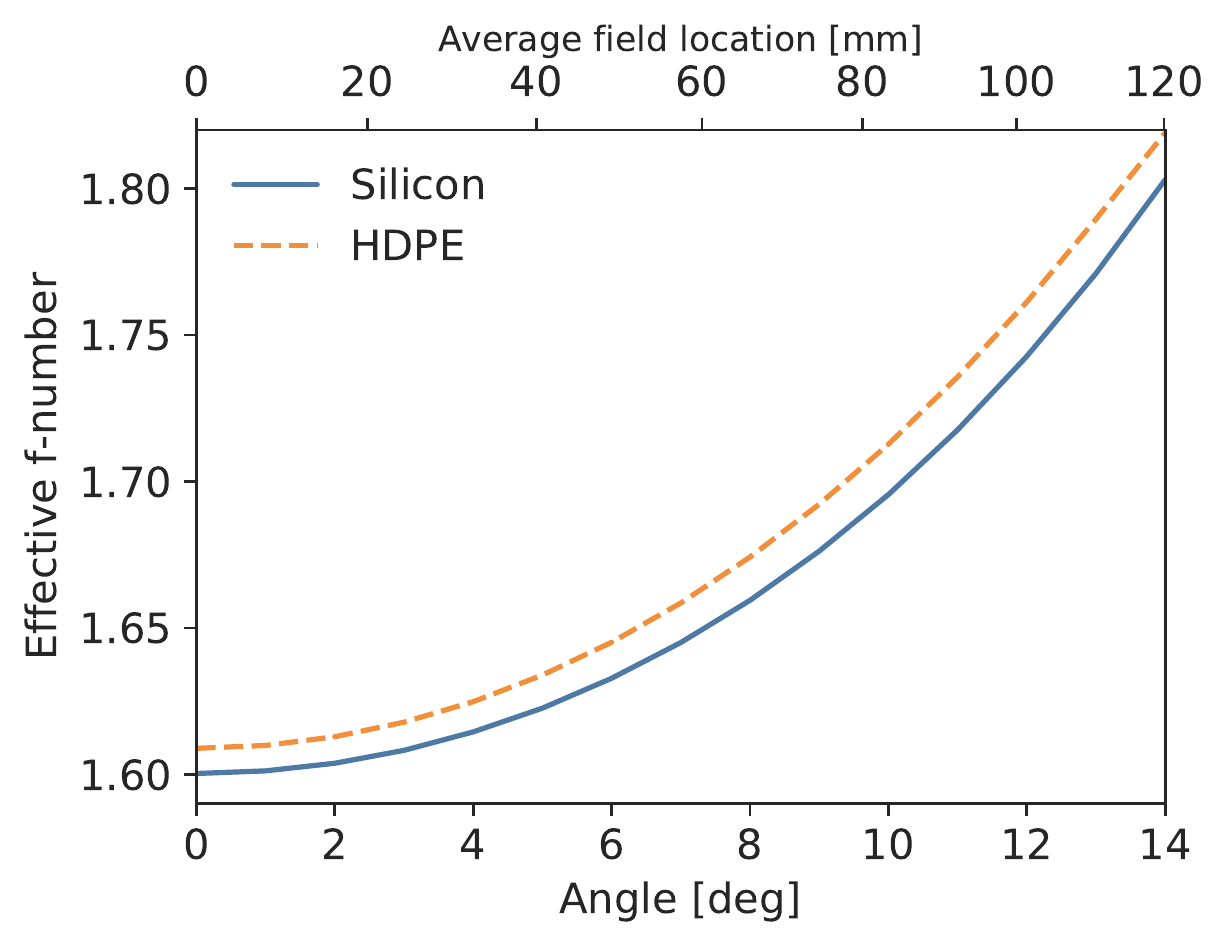}
\caption{The effective $f$-number across the field of view. The solid blue and dashed orange lines correspond to the silicon and HDPE designs, respectively. The upper horizontal axis shows average location on the focal plane while the lower horizontal axis shows the corresponding angle on the sky. Note that since the two telescopes have slightly different plate scales, the field location corresponds to the average of the two designs.}
\label{fig:fnumber}
\end{center}
\end{figure}

Figure \ref{fig:fnumber} shows the effective $f$-number across the field of view,
\begin{equation}
F = \frac{1}{2NA},
\end{equation} 
where $NA$ is the numerical aperture calculated through an integral across the telescope stop based on information from tens of thousands of ray tracing results (see definition for $NA$ in \cite{Zemax, Siew2005, Born1999}). The on-axis $f$-numbers are $f$/1.60 and $f$/1.61 for the silicon and HDPE designs, respectively. Both numbers grow at essentially the same rate with increasing field location. The effective $f$-number for both designs is approximately $f$/1.8 at the edge of the field (14 deg). This has implications for the amount of power that spills past the aperture stop (see Sec \ref{sec:po}\ref{sec:et}).

\subsection{Field of view and vignetting}

Although we limit the discussion in this paper to a 28~degree field of view, the silicon design should be able to support an even wider diffraction-limited field of view. At 16 and 17~degrees off axis, however, Zemax indicates that approximately 7 and 23\% of the rays emitted from the focal plane are vignetted, respectively. At a 17-degree angle, the 350-GHz Strehl ratio is 0.87. Since vignetting impacts the image illumination and therefore the thermal loading on CMB experiments, we argue that the silicon design can realistically support up to about a 32-deg diffraction-limited FOV without being significantly vignetted. In contrast, the HDPE design starts to vignet extensively at field angles exceeding 14 degrees while the Strehl ratio also falls off rapidly at larger angles. The HDPE design cannot support a larger field of view.

\subsection{Tolerancing}
\label{sec:tolerancing}

Tolerance against mechanical misalignment and variation in material properties is an important aspect of any optical design. The two systems being considered perform somewhat differently under a simple tolerancing analysis; an effect that can be attributed to the larger baseline RMS wavefront error of the HDPE design compared to the silicon design. We limit our study to variation in: radii of curvature, translations, rotations, and lens index of refraction. Among these parameters, the tolerancing analysis suggests two design aspects of primary importance: radius of curvature and index of refraction.

Due to its high index of refraction, the silicon design demonstrates a strong dependence on the radius of curvature; a 1\%-variation in the radius of curvature reduces the average (across the field of view) Strehl ratio down to 0.80 at 350 GHz. Conversely, the HDPE design is less sensitive to percent-level errors in radius of curvature. Instead, the plastic design is particularly sensitive to errors in the index of refraction.

\begin{figure}
\begin{center}
\includegraphics[width=9cm]{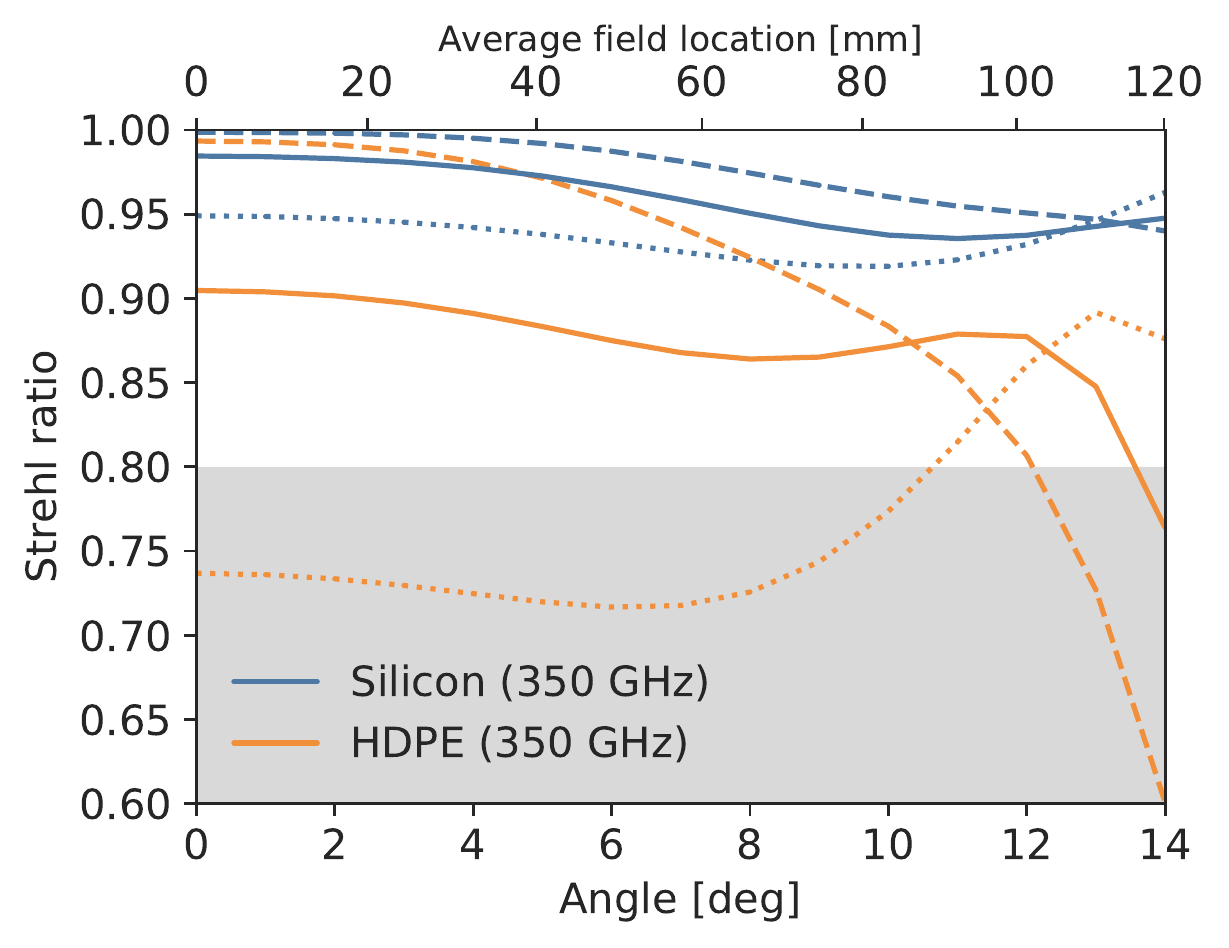}
\caption{Strehl ratio at 350 GHz as a function of field angle for the two designs. Solid lines correspond to the Strehl ratio if the lens index of refraction agrees with the design values, whereas dotted and dashed lines correspond to a scenario where the index is 0.2\% lower and greater than the nominal value, respectively. It is clear that the HDPE design is quite susceptible to variation in the index of refraction.}
\label{fig:strehl_ratio_tolerancing}
\end{center}
\end{figure}

Figure \ref{fig:strehl_ratio_tolerancing} shows the 350-GHz Strehl ratio across the field of view if the lens material index of refraction differs from the design value at the 0.2\% level. It is clear that the optical performance of the HDPE telescope depends critically on the index of refraction. Unfortunately, the literature on HDPE and silicon index of refraction at 4-K temperatures is somewhat limited \cite{Lamb1996} and variations between material suppliers are known to be substantial. The two designs presented here would benefit significantly from an improved constraint on the cryogenic index of refraction of these materials.

\section{Physical Optics}
\label{sec:po}

We use a \textit{Python} API to effectively interface with GRASP and run a large number of simulations of the far field beam response for the two proposed designs \cite{Prather1997, Sorensen2007, GRASP2018}.\footnote{GRASP is an antenna and optical modelling software capable of providing physical optics and method of moments calculations at mm-wavelengths. See: \href{https://www.ticra.com/}{https://www.ticra.com/}} All physical optics (PO) simulations are performed in the time-reverse sense (transmit mode), with an electric field emitted by a pixel on the focal plane and then propagated in succession through optical elements.\footnote{In GRASP, the PO simulations are run for cases where the lenses are mounted in opaque screens \cite{GRASP2018}.} The far field beam response is found by integrating over the equivalent surface current distribution of the final optical element (the primary lens). 

We perform simulations for pixels along three directions at the focal plane which correspond to 0, 45, and 90 degrees azimuthal angle (see red, green, and blue lines in Figure \ref{fig:distribution}), labelled X, D, and Y from now on. The polarized pixel beam used to illuminate the lenses outputs an electric field that oscillates along the X-axis (red line), corresponding to 0-deg azimuthal angle. The sampling of the three azimuthal directions allows us to probe polarization dependence in the predicted beam response. The alignment of the detector polarization vector affects both beam ellipticity and cross polarization response (see Sections \ref{sec:po}\ref{sec:ellipticity} and \ref{sec:po}\ref{sec:cross_pol}).

The physical optics simulations presented in this section ignore both internal reflections and reactive interactions with passive optical elements such as internal baffles. This is a significant approximation of a real optical system. Under this approximation, only the forward-propagating power (in time reverse sense) is used to estimate the far field beam. To an extent, this would be comparable to a system with both perfect internal baffling and perfect anti-reflection coating across all frequencies and incidence angles. This aspect of the simulations has the effect of underestimating asymmetric beam power since internal reflections that eventually make it out to the sky have not propagated along the expected optical path. This also implies that the total beam solid angle is underestimated in all results presented in this section. A more advanced analysis would incorporate full-wave solutions such as method of moments, or a double PO approach \cite{Sorensen2014}, but we leave this for later work, in part because of the associated computational challenges.

Because of these approximating assumptions and the simple nature of the optical system, the PO simulations are sufficiently fast that they can be generated for hundreds of mm-wavelength detectors in a reasonable amount of time (few days) on a workstation computer.\footnote{Analysis computer equipped with two Intel Xeon E5-2697v4 18-core processors and 512 GB of RAM.} The resultant electric field can then be decomposed along an arbitrary vector basis in both the near and the far-field of the instrument. 

\begin{figure} 
\begin{center}
\includegraphics[width=9cm]{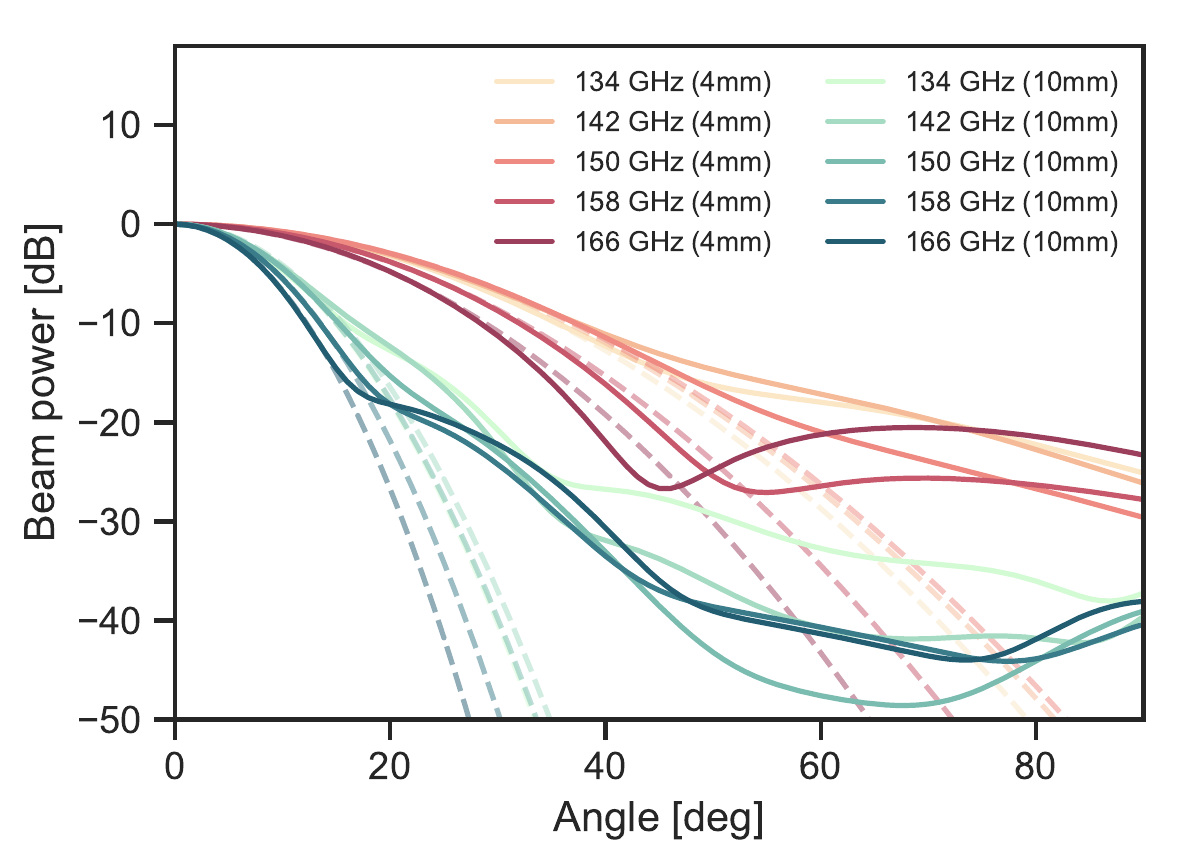}
\caption{Some of the monochromatic pixel beams that are used as input for the 150-GHz band-averaged results. Dashed lines correspond to the best-fit Gaussian equivalents. It is clear that the assumed pixel beams have considerably more solid angle than the Gaussian equivalents. On average, we find that the 4- and 10-mm pixel beams have 8 and 14\% more solid angle than their Gaussian approximations, respectively.}
\label{fig:pixel_beams}
\end{center}
\end{figure}

\subsection{Pixel beams}
\label{sec:pixel_beams}

The choice of pixel beam input critically impact these physical optics simulations. We study a range of pixel apertures to understand how the stop edge taper impacts both the far field beam response and stop spillover. Instead of assuming a Gaussian pixel beam model, we base our pixel beams on HFSS simulations for a photolithographed bolometer array coupled to spline-profiled feedhorns, similar to those designed by for ACTPol and Advanced ACTPol \citep{Niemack2010, Koopman2016, Simon2016}.\footnote{Ansys HFSS is an electromagnetic field simulation software that is frequently used to simulate the performance of high-frequency electronics products such as RF antennas. \href{https://www.ansys.com/products/electronics/ansys-hfss}{https://www.ansys.com/products/electronics/ansys-hfss}} Figure \ref{fig:pixel_beams} shows some of the beam profiles of the 150-GHz pixel beams assuming 4- and 12-mm pixel apertures. 

Pixel beam illumination patterns critically impact the far field beam \cite{Goldsmith1998, Goldsmith1999, Griffin2002}. In the time reverse sense, small pixel apertures illuminate the stop quite evenly which maximizes the beam FWHM and symmetrizes the beam. However, a small pixel aperture also leads to increased cold stop spillover and diffraction effects. Realistic pixel beam simulations commonly predict significant ($> 1\%$) sidelobe power compared to the best-fit Gaussian model, especially for devices with non-negligible bandwidths. Figure \ref{fig:pixel_beams} shows that we observe substantial difference in the beam solid angle between the pixel beams used in these simulations and the corresponding best fit Gaussian models. The fact that the Gaussian pixel beam model tends to underestimate the solid angle can obviously impact both detector loading and far field beam estimates. 

Although these simulations incorporate pixel beams that are significantly more realistic than a simple Gaussian beam, it is important to note that the pixel beam profiles have not been optimized for the two telescope designs presented in this work. This will impact spillover and far field beam response to some extent. However, we argue that since we are trying to understand general performance of the designs as a function of stop illumination, which obviously scales with pixel aperture size, careful tuning of the pixel beam is not critical.

For discussion in subsequent sections, we will present analysis where the pixel apertures are varied over 4-10 mm in 1-mm increments. For the 90 GHz band, the 4, 5, 6, 7, 8, 9, 10-mm pixel apertures correspond to roughly --3.4, --4.8, --6.5, --8.8, -11.2, -13.9, and --16.7 dB center pixel edge tapers in the HDPE design. The corresponding edge tapers for the silicon design are --3.5, --5.2, --7.2, --9.9, --12.6, --15.6, and --18.8 dB. We note that the effective edge taper varies across the field of view (see Section \ref{sec:po}\ref{sec:et}). In order to simulate the finite bandwidth of CMB bolometers, typically about 20--30\%, we run all physical optics simulations at five different evenly weighted frequencies within a 25\% wide frequency band centered on the main frequency (90, 150, and 280 GHz). For all simulations, the far field beam pattern of the input pixel beams are azimuthally symmetric and have zero cross-polar terms. We realize that this is not representative of real horn antennas, which in particular will admit cross-polar modes at some level. However, we choose not to incorporate pixel cross-polarization so that geometrical- and polarization-dependent effects captured in these physical optics simulations are cleanly distinguished from pixel-related effects. 

\begin{figure}
\begin{center}
\includegraphics[width=9cm]{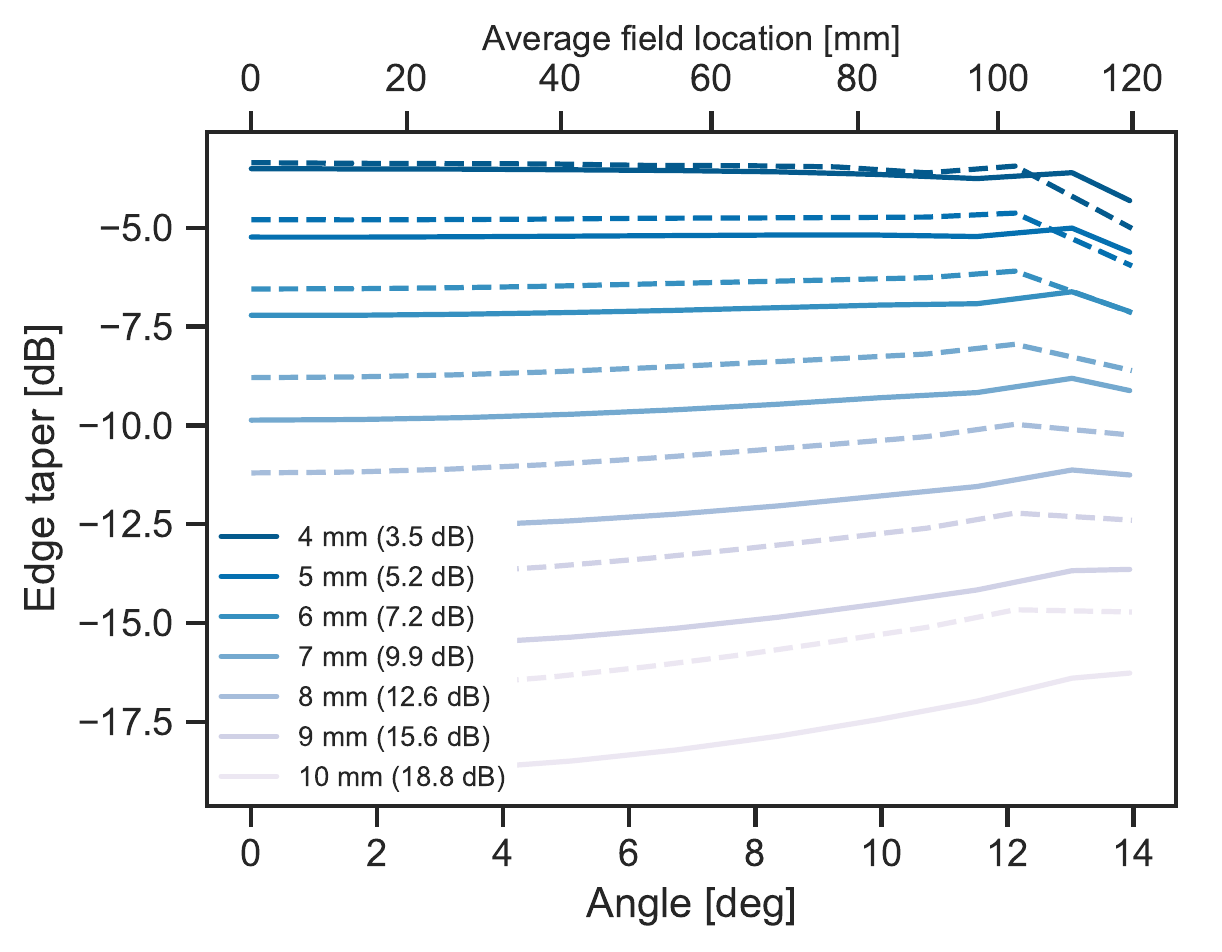}
\caption{The edge taper as defined by Equation \ref{eq:et} for 90-GHz pixels of different aperture sizes. The solid lines correspond to the silicon design and the dashed lines are the HDPE design. The fall-off at high field locations is related to the relative increase in secondary lens spillover which complicates the interpretation of edge taper as defined by Equation \ref{eq:et}. The legend indicate the pixel aperture size and the corresponding on-axis edge taper calculated according to Equation \ref{eq:et}.}
\label{fig:et}
\end{center}
\end{figure}

\subsection{Edge taper and cold spillover}
\label{sec:et}
The edge taper is typically defined as the beam power density at the edge of the aperture stop relative to the peak. In simple on-axis refractor systems, pixels with azimuthally symmetric beam patterns placed at the center of a focal plane have unambiguously defined edge tapers. For off-axis pixels, we can define the edge taper as the azimuthally averaged power along the stop edge
\begin{equation}
T_\mrm{e} =  \frac{1}{2\pi} \int _S d \phi P(\theta_\mrm{edge}, \phi) / \max (P(\theta, \phi) )
\label{eq:et}
\end{equation}
where $P(\theta_\mrm{edge}, \phi)$ is the beam power along the aperture rim. This expression clearly reduces to the traditional definition for on axis system
\begin{equation}
T_\mrm{e} =  P(\theta_\mrm{edge}) / P_0.
\end{equation}
where $P_0$ is the peak beam power.

Using a simple Gaussian beam formalism for an on-axis beam \cite{Goldsmith1998, Padin2010}, we can see that an increase in $f$-number, $F$, leads to more aggressive edge tapers. Starting by writing the electric field distribution on the stop as
\begin{equation}
E_s(r) = \exp \left[  -\left( c \frac{r}{a_s}\frac{w\pi}{F\lambda} \right)^2 \right],
\label{eq:gauss_form}
\end{equation}
where $a_s$ is the stop radius, $r$ is the radial coordinate, $w$ is the Gaussian beam waist, $\lambda$ is the wavelength, and $c$ is a constant, it becomes clear that
\begin{equation}
T_\mrm{e} =  \exp \left[  -2\left( c \frac{w\pi}{F\lambda} \right)^2 \right].
\end{equation}
This then implies that
\begin{equation}
F_\mrm{e} = 1 - T_\mrm{e},
\end{equation}
where $F_\mrm{e}$ the represents the fractional amount of power that makes it through the stop. This expression, however, does not apply when we have defined the azimuthally averaged edge according to Equation \ref{eq:et}.

In general, the above formalism can easily mask subtleties associated with off axis and complicated stop illumination patterns. It holds that spillover efficiency can be calculated according to
\begin{equation}
\eta _s = \frac{\int _{\mrm{stop}} | E_s |^2 d\bf{S}}{\int _{\mrm{\infty}} | E_s |^2 d \bf{S}},
\label{eq:etas}
\end{equation}
and the above expression can be numerically integrated for arbitrary stop illumination patterns. We use this numerical approach for all subsequent analysis.

Asymmetric stop illumination patterns become particularly relevant for detector array telescopes designed to push the limits of optical throughput, such as the ones mentioned in Section \ref{sec:throughput}, which often have edge tapers and $f$-numbers that vary significantly across the field of view. Some care is therefore warranted when using Gaussian beam formalism to connect $f$-number statistics derived from radiometric (geometrical) illumination \cite{Siew2005} to cold stop spillover.

Figure \ref{fig:et} shows the 90-GHz edge taper at the aperture stop for a set of pixel apertures ranging from 4-10~mm for both silicon and HDPE design. Depending on the pixel aperture and design, we observe about a 1--4 dB variation in edge taper across the field of view. As expected, this variation is strongest for the largest pixel apertures. Our extension of the edge taper definitions starts to break down at the edge of the field where a significant amount of beam power is spilled past the secondary lens. This can be seen as a drop in the edge taper for large field angles. The difference in 90 and 150 GHz edge taper for a fixed pixel aperture is quite strong. This has implications for multichroic detector architectures \cite{Suzuki2012, Thornton2016, Anderson2018, Galitzki2018, Matsumura2016}.

\begin{figure*}[t]
\begin{center}
\includegraphics[width=13cm]{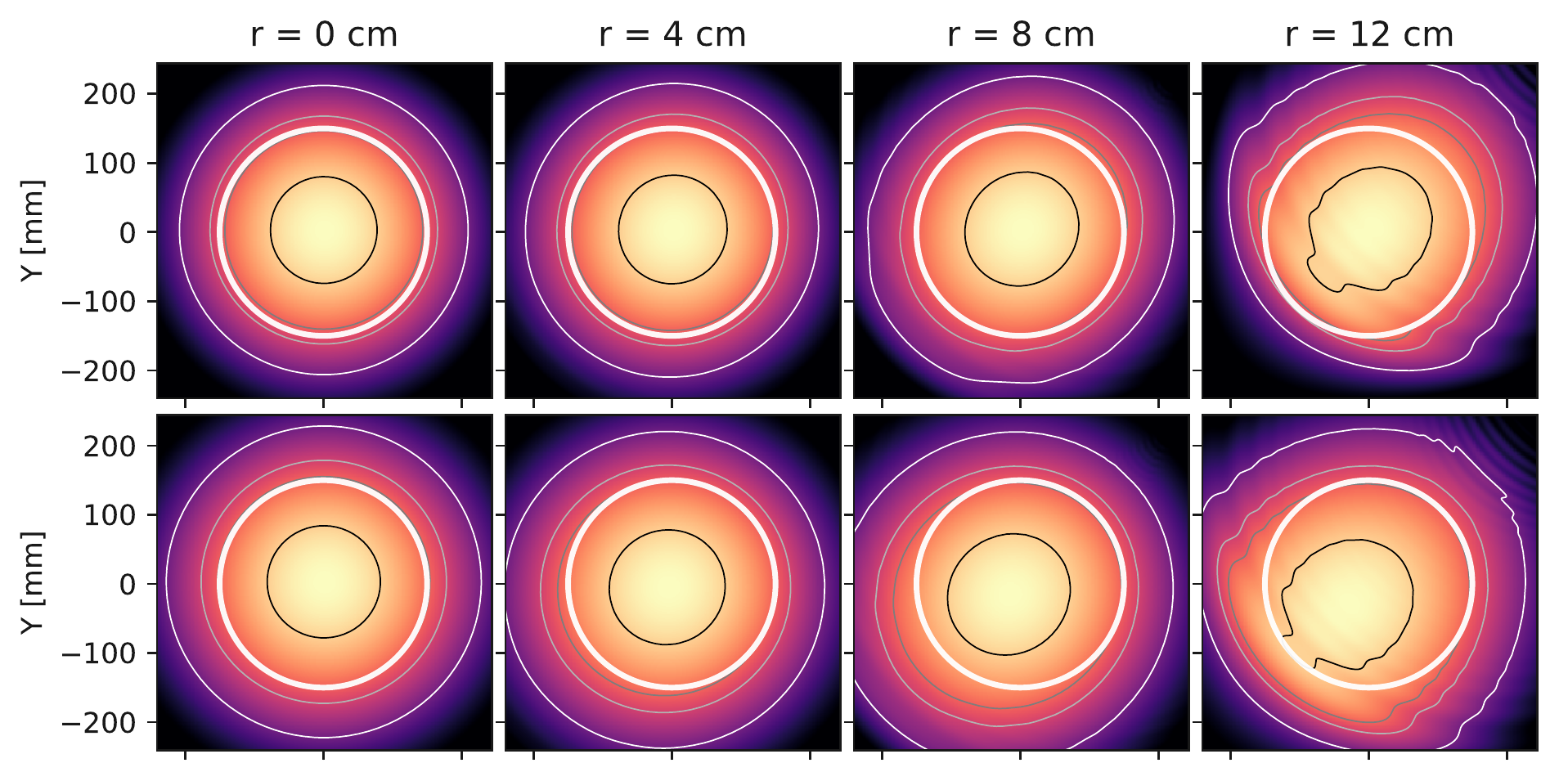}
\caption{The total beam power calculated on a Cartesian coordinate system coinciding with the aperture stop (the inner surface of the primary lens) for a 5-mm pixel aperture at 150 GHz. The top and bottom rows correspond to the silicon and HDPE designs, respectively. The four columns represent four different field location with $r=12$~cm corresponding to the edge of the field. The beam has been normalized and the contour lines correspond to --3, --10, --13, and --20~dB. The colorscale spans --30-0 dB and each panel is 450 mm across. The thick white circle represents the aperture stop (300 mm diameter). These spillover maps were generated by combining results at five different frequencies within the frequency band. The interference patterns seen for the pixels at the edge of the field become much more pronounced for single-frequency beam maps. The distorted power distributions at the edge of the field, $r=8$ and $r=12$~cm are in part due to vignetting of the lenses.}
\label{fig:spill}
\end{center}
\end{figure*}

The physical optics simulations allow us to calculate the field at an arbitrary location inside the telescope. Figure \ref{fig:spill} shows the power projected on the cold stop for a few 150-GHz 5-mm aperture detectors for the two designs. As expected, the stop illumination becomes less uniform near the edge of the field. Interestingly, we see that the stop power distributions are translated significantly relative to the stop center for the HDPE design; this is a direct consequence of the reduced telecentricity.

When calculating total cold spillover, we combine the spillover past the secondary lens to the spillover past the cold stop (primary lens back surface). The spillover past the secondary lens is calculated according to
\begin{equation}
\eta _{\mrm{L}_2} = 1 - \frac{\int _{\mrm{L}_2} | E_{\mrm{L}_2} |^2 d\bf{S}}{\int _{\mrm{\infty}} | E_{\mrm{L}_2} |^2 d\bf{S}}
\end{equation}
where $|E_{\mrm{L}_2}|^2$ is the total (squared modulus) electric field distribution emitted by the pixel beam. Since the cold stop coincides with the inside of the primary lens, we can define the effective cold stop (primary lens) spillover as
\begin{equation}
\eta _{\mrm{L}_1} = (1 - \eta _{\mrm{L}_2}) \times \left( 1 - \frac{\int _{\mrm{L}_1} | E_{\mrm{L}_1} |^2 d\bf{S}}{\int _{\mrm{\infty}} | E_{\mrm{L}_1} |^2 d\bf{S}}  \right) 
\end{equation}

where $L_1$ and $L_2$ indicate the primary and secondary lenses respectively. The effective cold stop spillover, $\eta _{\mrm{L}_1}$, represents the amount of power that doesn't make it through the primary lens.

\begin{figure}[t]
\begin{center}
\includegraphics[width=9cm]{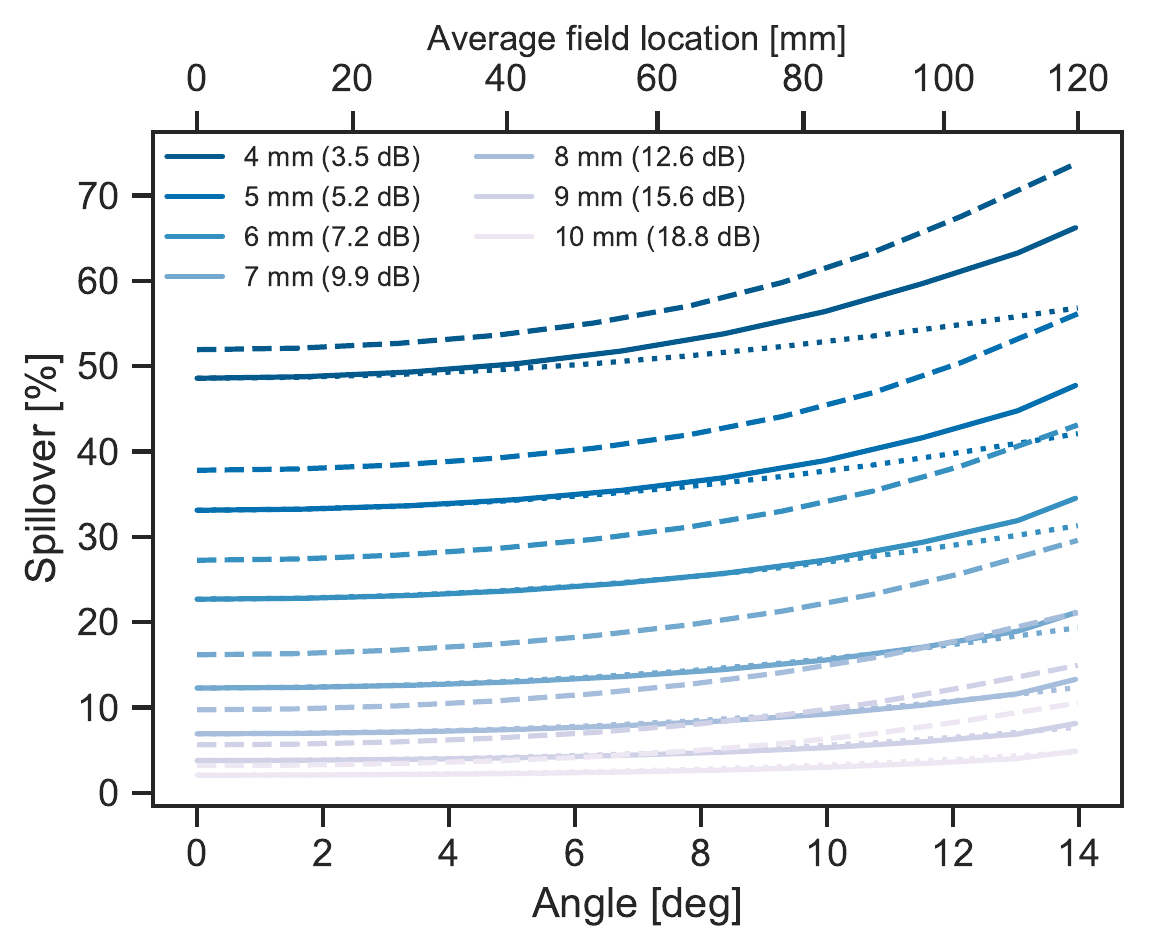}
\caption{The 90-GHz cold stop spillover as a function of field location for the different simulated pixel apertures. Solid and dashed lines correspond to the silicon and HDPE designs, respectively. Dotted line corresponds to spillover as determined from the Gaussian beam formalism (see Equations \ref{eq:gauss_form} and \ref{eq:etas}) after normalizing to the silicon beam spillover at the center of the field. The legend indicate the pixel aperture size and the corresponding on-axis edge taper calculated according to Equation \ref{eq:et}.}
\label{fig:spill_dist}
\end{center}
\end{figure}

Figure \ref{fig:spill_dist} shows the total cold spillover as a function of field location for different pixel apertures.\footnote{As stated earlier, the physical optics simulations ignore all internal reflections in the system and treat the side walls of the optics sleeve as perfect absorbers at all incident angles and frequencies. We expect that reflections from the side walls will reduce total spillover somewhat but also cause a general increase in systematic effects.} The solid and dashed lines correspond to the silicon and HDPE designs respectively. The dotted line represents an $f$-number scaling obtained from the Gaussian beam formalism described by Equations \ref{eq:gauss_form} and \ref{eq:etas} normalized to coincide with the spillover for the silicon design at the center of the focal plane. Judging by the discrepancy between the solid and dotted lines, we can surmise that the $f$-number scaling that is determined from a simple Gaussian beam formalism is not always accurate. We note for example a roughly 10\% difference at the edge of the field for small pixel apertures. This is particularly relevant for systems where internal optical loading represents a significant fraction of the total detector noise budget. In those cases, spillover and mapping speed calculations should be performed using this type of analysis for improved accuracy. The need for this approach becomes even more apparent when less directive pixel beams are used; this includes pixel technologies deployed on SPIDER, the BICEP and Keck Array experiments, and Polarbear \cite{Obrient2015, Suzuki2012}. It is worth noting that the Gaussian beam formalism in good agreement with the silicon reference case for compact pixel beams (low edge taper), but starts to diverge when we decrease the effective pixel size.

\subsection{Ellipticities}
\label{sec:ellipticity}

The beams produced by the simulations presented in this paper are in all cases quite well approximated by an elliptical Gaussian beam model in the vicinity of the peak response. Beam ellipticity characterizes first order deviations from azimuthal symmetry of far field beam response. The ellipticity can be defined as 
\begin{equation}
e = \frac{\sigma_{\mrm{max}}-\sigma_{\mrm{min}}}{\sigma_{\mrm{max}}+\sigma_{\mrm{min}}},
\end{equation}
where $\sigma _{\mrm{max}}$ and $\sigma _{\mrm{min}}$ correspond to the widths of the best-fit Gaussian envelope to the semi-major and semi-minor axis of the co-polarized far field beam response. Beam asymmetries generate spurious polarization signals which can mimic an actual sky signal and bias science results \cite{Souradeep2001, Rosset2007, Keihanen_2012, Bicep2Keck_IV_2015, Shimon2008}. 

We fit a six-parameter elliptical Gaussian in an approximately 2-deg diameter region centered on the main beam. The fitting function uses \texttt{scipy.optimize.leastsq} on a beam map that is sampled with a resolution of 14.3 arcsec. The results are slightly sensitive to the region and resolution used for the fit, but we find that our results are essentially unaffected by reasonable variations in both resolution and main beam coverage.

\begin{figure}[t]
\begin{center}
\includegraphics[width=9cm]{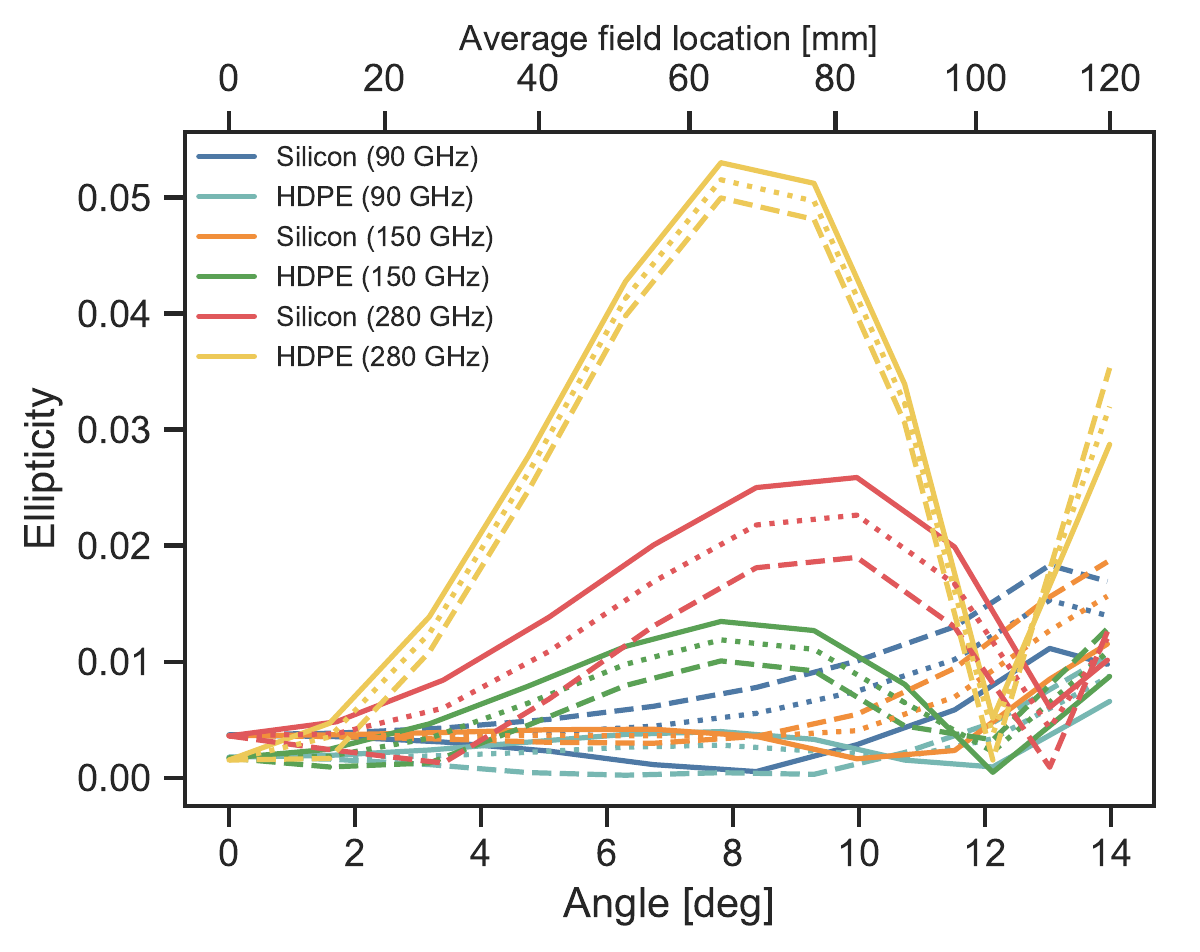}
\caption{Beam ellipticities as function of field location for the three simulated frequencies (90, 150, and 280 GHz) and the two telescope designs. There are three types of curves for every color: the solid, dashed, and dotted lines correspond to cases where the pixel is translated along the X-, D-, and Y-direction on the focal plane, respectively. In all cases, the electric field emitted by the pixel is oscillating along the X-direction. Somewhat unsurprisingly, the predicted ellipticity depends on the alignment of the pixel beam polarization with the projected shape of the stop. It is interesting to see a bump in the predicted ellipticity for the 280-GHz silicon design at intermediate field values. At the edge of the field, the ellipticity starts to rise sharply again.}
\label{fig:ellipticities}
\end{center}
\end{figure}

Figure \ref{fig:ellipticities} shows the beam ellipticities across the field of view for the two telescope models assuming 8, 6, and 4 mm pixel apertures for the 90, 150, and 280 GHz bands, respectively.\footnote{Note that the best-fit beam ellipticity depend slightly on the data that are used to determine the fit. We fit an elliptical Gaussian model in circular region roughly 3~FWHM in diameter centered around the beam center.} Note that the three pixel apertures, 8, 6, and 4 mm for the three frequencies correspond to approximately -11.2, -7.2, and -3.5~dB center pixel edge tapers, respectively. We perform these simulations along three axis on the focal plane (see Figure \ref{fig:distribution}) and see a clear difference in the amount of ellipticity depending on the alignment of the detector polarization angle. This is understandable given the electric field distribution as projected on the stop varies with detector polarization. It is interesting to note a turnaround point at intermediate field locations for some of the curves. This is caused by an interplay between optical aberration and secondary lens spillover, the latter of which is not propagated in these simulations.

At 280 GHz, the average ellipticity of the HDPE design appears to be significantly larger than that of the silicon design. In contrast, the dispersion between polarization axes is larger for the silicon design. This is likely due to the higher index of refraction of silicon which increases polarization dependence (see Section \ref{sec:po}\ref{sec:cross_pol}). Simulations presented in \cite{Duivenvoorden2018} suggest that systematics due to the beam ellipticities of these two designs would not have a significant impact on the science goals of a 4th-generation CMB satellite experiment searching for a primordial $B$-mode polarization signal.

\begin{figure}[t]
\begin{center}
\includegraphics[width=9cm]{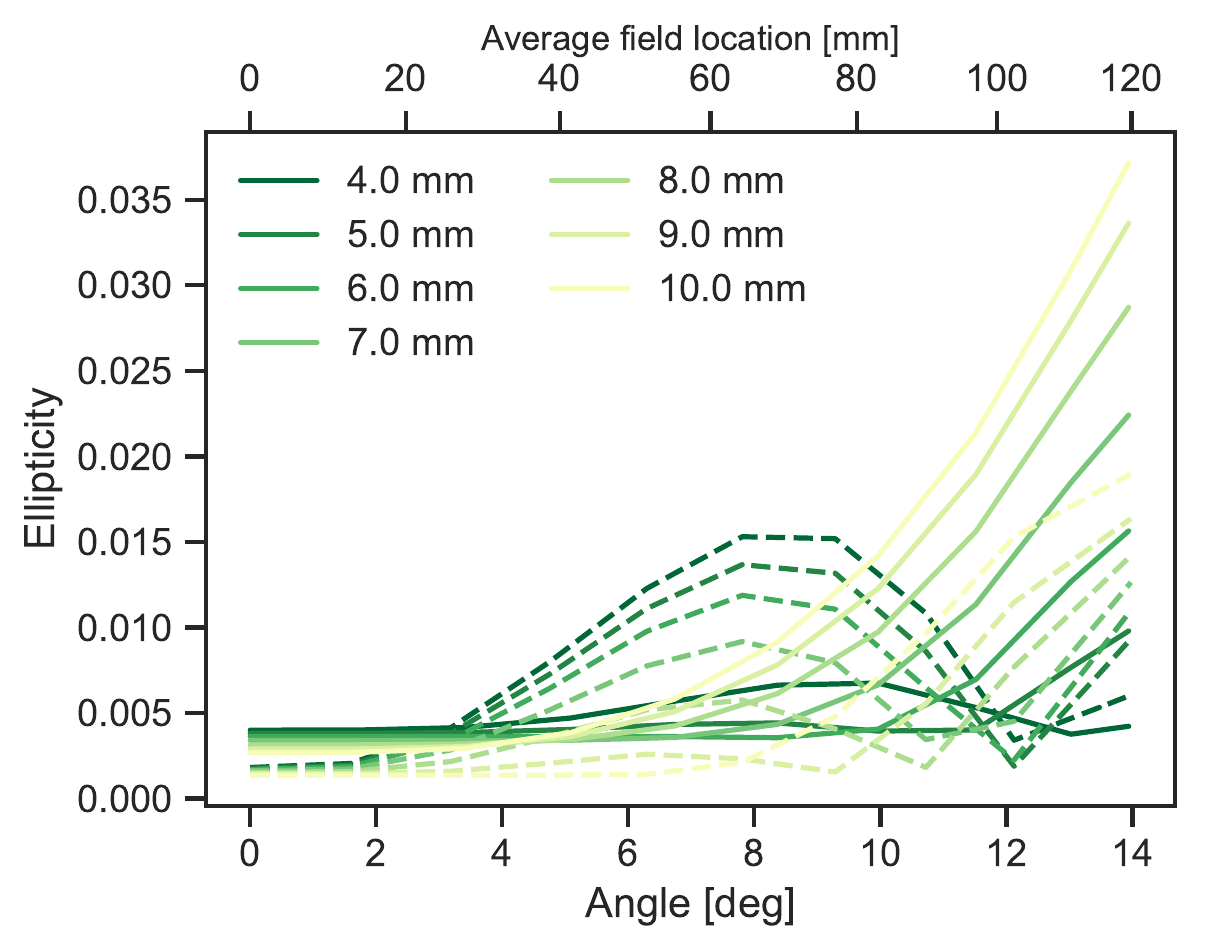}
\caption{Beam ellipticity at 150 GHz as a function of field location along the D-direction for different pixel apertures ranging from 4 to 10~mm. Solid lines correspond to silicon design while dashed lines are the HDPE design. It is interesting to see such significant difference in beam ellipticity for the two designs.} 
\label{fig:psize_ellipticities}
\end{center}
\end{figure}

We also study the dependence of beam ellipticity on the input pixel beam aperture. Figure \ref{fig:psize_ellipticities} shows the beam ellipticity along the focal plane diagonal for five different pixel apertures ranging from 4 to 10 mm. It is well known that far field beam solid angle depends on the stop illumination pattern and therefore the pixel beam aperture \cite{Griffin2002}. However, here we also demonstrate how the far field beam ellipticity depends on the pixel aperture as well as the overall optical architecture. Note that, by default, the Strehl ratio as calculated using Zemax is insensitive to pixel aperture size and the edge taper.

\subsection{Beam solid angle}
\label{sec:bsa}
The co-polar beam solid angle is defined as
\begin{equation}
\Omega _\mrm{beam} = \int d\Omega (P_\mrm{co}(\theta,\phi) + P_\mrm{cross}(\theta,\phi)) / \max{ \left( P_\mrm{co}(\theta,\phi) \right)},
\end{equation}
where $P_\mrm{co}(\theta, \phi)$ and $P_\mrm{cross}(\theta, \phi)$ are the co-and cross-polar beam response (antenna gain) as a function of sky location. For a perfect Gaussian beam we can show that the beam solid angle is $\Omega _\mrm{G} = 2\pi \sigma ^2$, where $\sigma$ corresponds to the Gaussian beam width. A diverse collection of optical non-idealities, including basic diffraction and scattering mechanisms, tends to increase the nominal solid angle of the beam. The variation in far field beam solid angle across the field of view provides direct indication of the aperture stop illumination non-uniformity.

\begin{figure}[t]
\begin{center}
\includegraphics[width=9cm]{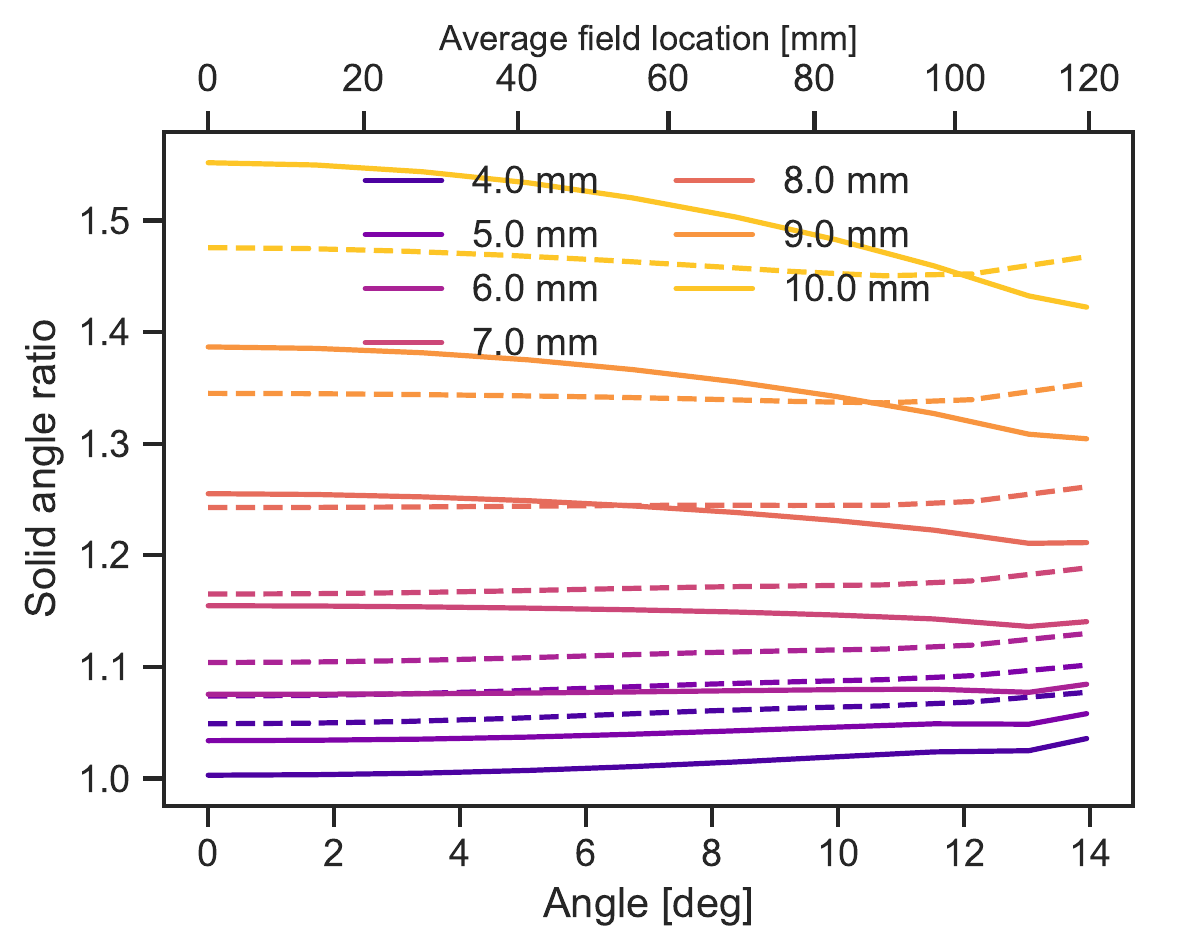}
\caption{Beam solid angle at 90 GHz as a function of field location for different pixel apertures ranging from 4--10~mm. Solid lines correspond to silicon design while dashed lines are the HDPE design. Small pixels give uniform stop illumination and far field beam sizes that approach the theoretical minimum. A negative slope is an artifact of secondary spillover which is ignored in the physical optics simulations.}
\label{fig:psize_bsa}
\end{center}
\end{figure}

Figure \ref{fig:psize_bsa} shows the corresponding 90-GHz beam solid angle variation across the field for five different pixel apertures. As expected, the smallest pixels illuminate the aperture stop most uniformly and therefore have the most compact far field beams. The beam solid angles are normalized by 
\begin{equation}
\Omega _\mrm{ideal} = 2 \pi \sigma _\mrm{ideal} ^2
\end{equation}
where,
\begin{equation}
\sigma _\mrm{ideal} =  \frac{1}{\sqrt{8\ln{(2)} }} \left( \frac{ \lambda }{D} \right),
\end{equation}
with $D = 30$~cm. For an 8-mm pixel we see that the beam solid angle is roughly 20-30\% greater than the theoretical limit. It is important to note that non-idealities such as internal reflections from the optics tube---which are not captured by these PO simulations---will most likely offset any negative trend in beam solid angle with field location.

\subsection{Cross polarization}
\label{sec:cross_pol}

We define cross-polarization amplitude as the ratio between the maximum in the co- and cross-polar beam responses 
\begin{equation}
x_\mrm{pol} = \max(P_\mrm{cross}(\theta,\phi)) / \max(P_\mrm{co}(\theta,\phi))
\end{equation}
where $P_\mrm{co}(\theta,\phi)$ and $P_\mrm{cross}(\theta,\phi)$ are the co- and cross-polar beam response as defined using Ludwig's 3rd definition \cite{Ludwig1973}. For an on-axis refractor illuminated by a linearly polarized pixel beam placed at the center of the focal plane, the far-field cross polarization pattern should be quadrupolar with nulls along the principal polarization axes and this is indeed what we observe in our simulations. The cross polarization depends heavily on the field location and is strongest for pixels placed along the diagonal field line. 

\begin{figure}[t]
\begin{center}
\includegraphics[width=9cm]{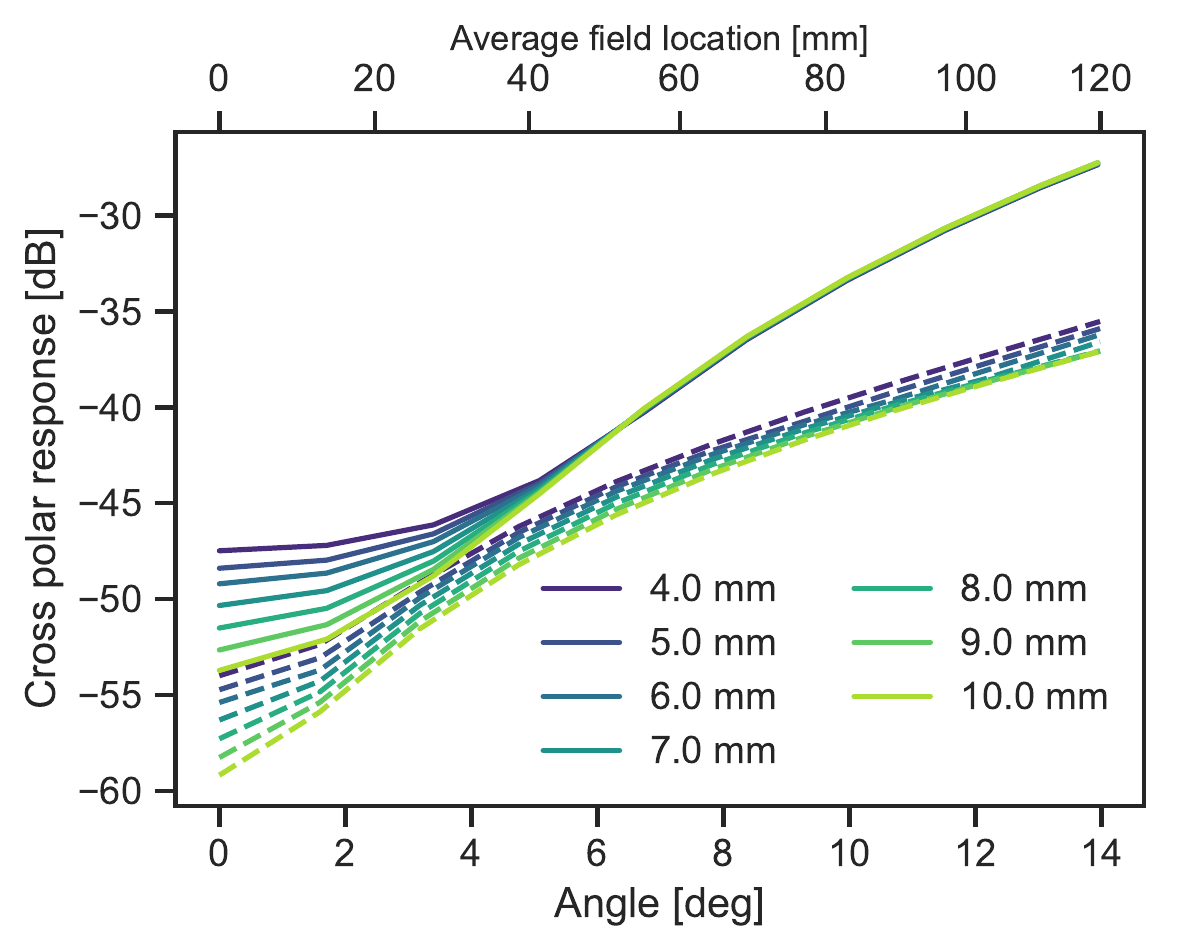}
\caption{The predicted 150-GHz cross-polar response for pixels along the D-direction. Solid lines correspond to the silicon design while dashed lines represent the HDPE design. We note a significant difference in the predicted cross polar response for the two designs.}
\label{fig:cross_polar}
\end{center}
\end{figure}

Figure \ref{fig:cross_polar} shows the predicted cross polar response as a function of field location along the diagonal for the two designs at 150 GHz. Note that the simulations do not predict significant frequency dependence in the cross polar response. It is interesting to see a significant difference in the cross polar response for the silicon and HDPE designs. At the edge of the field, the silicon design is predicted to have a --28~dB cross polar response while the corresponding value for the HDPE design is around --38~dB. This is caused by the high index of refraction of the silicon lenses which leads to larger polarization dependence in transmission. Since the physical optics simulations ignore both anti-reflection coatings and reflections, the difference in cross-polarization between the two refractors is maximal. By running method of moments simulations of a scaled version (20\% of original size) of the silicon refractors, we have found that a three-layer metamaterial AR coating will reduce the cross polarization, but still perform worse than the plastic lenses in general. Related to this topic, it is important to mention that for real experiments, the combination of antenna elements and bolometer crosstalk can be expected to cause cross polarization which dominates even the --28~dB level which is reported here \cite{Suzuki2012, Obrient2015, Simon2016}.

\subsection{Beam profiles}
\begin{figure}[t]
\begin{center}
\includegraphics[width=9cm]{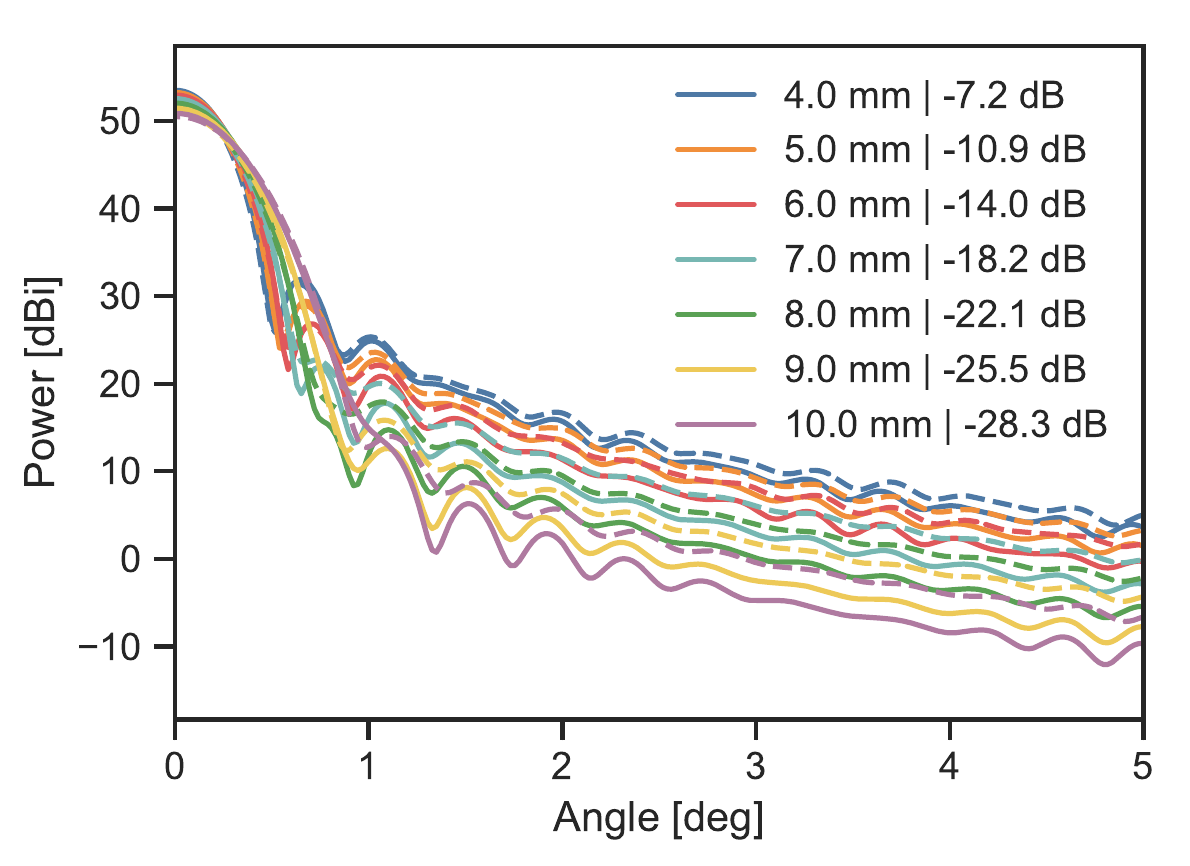}
\caption{Forward gain normalized beam profiles for a center pixel at 150 GHz. Different color curves correspond to pixel apertures ranging from 4 to 10 mm. The corresponding on-axis edge taper is also shown in the legend. Solid lines represent the silicon design while dashed line corresponds to HDPE.}
\label{fig:beam_profiles}
\end{center}
\end{figure}

\begin{figure}[t]
\begin{center}
\includegraphics[width=9cm]{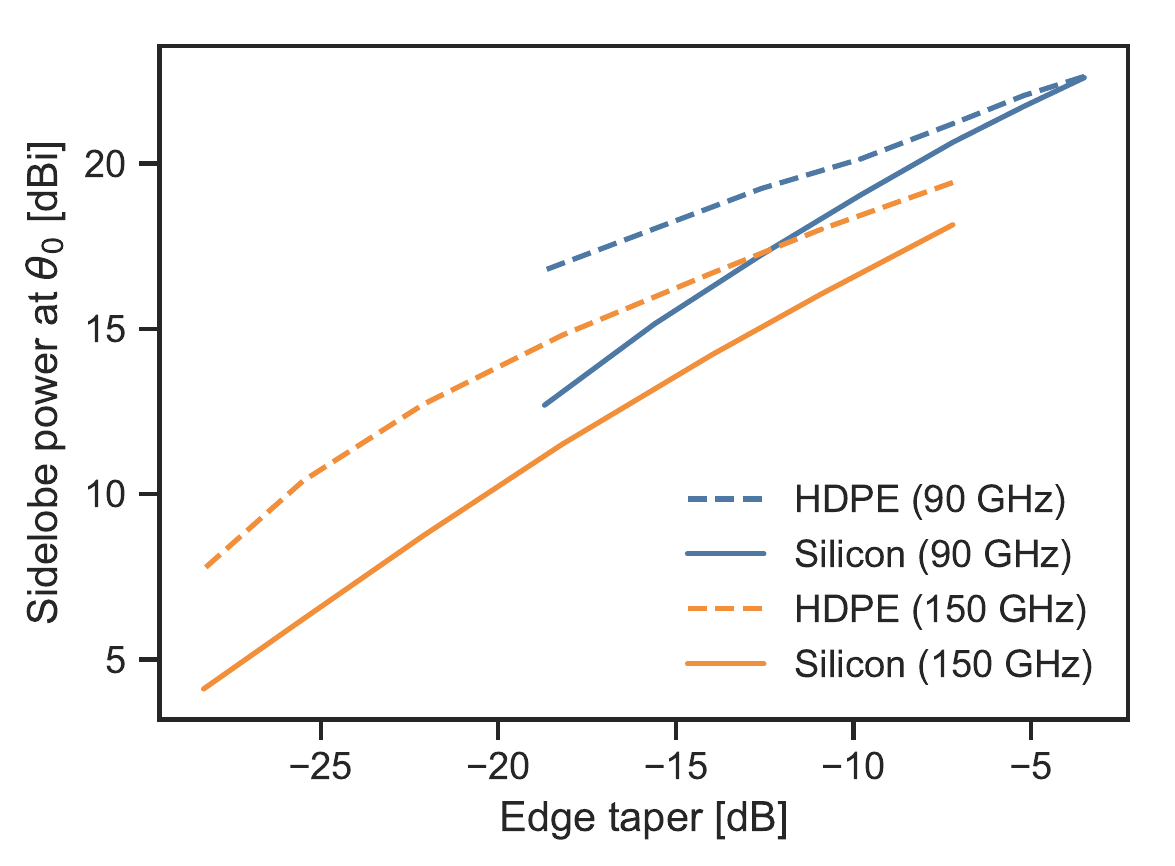}
\caption{Sidelobe amplitude at $\theta _0 = 1.5$~deg, as a function of edge taper. The relation for the two designs is shown at both 90 and 150 GHz.}
\label{fig:et_relation}
\end{center}
\end{figure}

It is instructive to look at the predicted co-polar far field beam profile for the two telescopes. Figure \ref{fig:beam_profiles} shows 150-GHz beam profiles for a center pixel for different pixel apertures (other pixels are qualitatively similar). The results follow expectations from basic antenna theory \cite{Kraus1986}. Apart from a clear difference in sidelobe amplitude as a function of edge taper, we observe relatively small differences between the two designs. The largest difference, approx 3-5 dB in the far sidelobe amplitude, is seen in the case of large pixel aperture (conservative edge taper). A simple model such as 
\begin{equation}
P(\theta) = A \sin ^{-3} (\theta),
\end{equation}
where $A$ is a constant, fits the beam profiles quite well in the region outside of the main beam. With this model, it is instructive to look at sidelobe amplitude, $A_0$, at some fixed angle, $\theta _0 = 1.5$~deg, as a function of edge taper for the two designs. Figure \ref{fig:et_relation} shows this relation for both 90 and 150 GHz. As expected, there is a clear power law relation between the sidelobe amplitude and the edge taper. Writing 
\begin{equation}
P ({\theta_0}) = k_0 S_e ^{\beta}, 
\end{equation}
which gives 
\begin{equation}
\log (P_{\theta_0}) = \beta \log(S_e) + \log(k_0),
\end{equation}
we find that $\beta$ = 0.38, 0.53, 0.65, 0.67 for HDPE at 90~GHz, HDPE at 150~GHz, silicon at 90~GHz, and Silicon at 150~GHz, respectively. This indicates that reducing the edge taper by 3~dB will drop the sidelobe amplitude at a given angle by 1.2--2.1 dBi. Obviously, such a relation is probably telescope dependent. 

\section{Comparison between physical and geometrical optics results}
\label{sec:gopocomp}
The physical optics calculations that are presented in this work are computationally intensive. It would be helpful to derive scaling relations that allow one to use geometrical optics results to make concrete predictions about the far field beam response. Unfortunately, the geometrical optics analysis presented in this work is insufficient for this. To better understand the relation between geometrical and physical optics results, one should investigate apodized (weighted) ray bundle inputs to the geometrical optics simulations; this is because the apodization process can mimic the aperture weighting set by the input pixel beam. With proper weighting, the time reverse ray tracing approach would likely produce Strehl ratios that correlated with non-ideality parameters predicted by physical optics. 

We leave comparison of apodized geometric optics and physical optics for later work. Despite this shortcoming, we argue that the analysis presented here shows how some geometrical performance metrics relate to properties such as cold stop spillover and far field beam response in the case of a simple refractor telescope. Although the qualitative results do not necessarily apply to all two-lens refractor designs, we hope that this work can provide guiding principles for telescope designs operating in the mm-wavelength; in particular the design of the CMB-S4 small aperture telescope design and the design of the optics for the LiteBIRD satellite \cite{S4_2016_Science_book, S4_2017_technology_book, Hazumi2019}.

\section{Discussion}
\label{sec:discussion} 

Future CMB observatories are aiming for unprecedented level of systematic control \cite{S4_2017_technology_book}. This calls for innovative optical designs informed by advanced simulation and analysis tools. The analysis presented in this paper demonstrate some benefits of an accelerated physical optics simulation framework that can quickly identify key design parameters and adapt to design changes. 

Unfortunately, this paper also suggests at least some level of disconnect between geometrical and physical optics simulations. Methods that connect these two approaches do exist \cite{Imada2016}, but they have not seen wide use in the design of wide field of view telescopes operating in the mm-range. At the level of accuracy needed for future CMB missions, both of these approaches might be inadequate, however. For example, the sidelobe profiles of the two telescopes presented in this paper vary significantly depending on the electrical properties of an encircling optics tube. Such interactions between passive and active optical components can only be realistically modeled using full-wave solutions (method of moments).

In general, even the simplest two-lens refractor systems suffer from uncertainties in the frequency response of detector beams, absorber properties, anti-reflection coating performance, material uniformity, machining tolerance, etc. Dedicated simulation and testing efforts are required to determine if these effects can impact the science goals of future CMB missions. It is also clear that advances in optical modeling need to be accompanied by further characterization of typical materials used in constructing telescopes operating at mm-wavelengths. For example, a detailed characterization campaign for basic lens materials such as silicon, HDPE, and alumina would be incredibly beneficial to the community. Such work should improve over measurements aggregated in \cite{Lamb1996} which only exist for a limited number of frequencies and temperatures.

Despite significant differences in Strehl ratios between the HDPE and silicon designs, we find that physical optics simulations predict quite comparable optical performance. For the design requirement used in this paper, one should therefore equally consider other performance metrics, such as mechanical stiffness, material uniformity, ease of manufacturing, weight, and thermal conductivity to influence design choices.

The results presented in Figure \ref{fig:cross_polar} are interesting, but maybe not surprising given the significant difference in the index of refraction for silicon relative to typical plastics. The results are consistent with basic expectations from the Fresnel equations. It is important to remember, however, that the physical optics simulations ignore any anti-reflection coatings and internal reflections that may or may not take place within and between lenses. To check if the trends seen in Figure \ref{fig:cross_polar} persist in more accurate full wave solutions, we have run method of moment calculations of $1/5$-scaled versions of the telescope designs with and without a three-layer AR coating. The anti-reflection coating is specifically designed to match the simulation frequencies. The results obtained from those simulations are qualitatively similar to Figure \ref{fig:cross_polar} even when incorporating a three-layer AR coating. The silicon design therefore appears to suffer from higher cross-polarization even when accounting for anti-reflection coating.

\section{Conclusions}
\label{sec:conclusions}

There is continued interest in the development of small aperture telescope designs for CMB observations. Throughout the design, build, and characterization process, decisions regarding cryogenic optics, $f$-numbers, temperatures of active and passive optical elements, baffling, etc., influence the sensitivity of an experiment. Some effort has been devoted to developing software to identify configurations that maximize mapping speed and minimize optical systematics. However, there is significant room for improvement and advances in optical modeling need to be accompanied by comprehensive testing and validation at both individual component and full system level.

In this paper, we have presented two refractor designs, one using silicon and another using HDPE lenses, that are both able to support 28-deg diffraction-limited field of view up to 280 GHz. We have compared their performance using both geometric and physical optics and shown that even though their Strehl ratios differ significantly, the far field beam properties predicted by physical optics are quite similar. The optical performance of the HDPE design starts to diverge from the silicon design at higher frequencies, such as 280 GHz. Arguably, the two designs differ most significantly in the tolerance requirements and the cross polarization performance, discussed in Sections \ref{sec:go}\ref{sec:tolerancing} and \ref{sec:po}\ref{sec:cross_pol}, respectively.

\section{Funding Information}
The author acknowledges support from the Swedish National Space Agency (SNSA/Rymdstyrelsen). 

\section{acknowledgments} 
I thank Marco De Petris, Simon Dicker, Adriaan J. Duivenvoorden, Hiroaki Imada, and Frederick Matsuda for many helpful comments on the manuscript. Also thanks to Patricio Gallardo, Johannes Hubmayr, Philip Mauskopf, Michael Niemack, Jeff McMahon, Sara Simon, and Lyman Page for informative discussions about optics and optics simulations over the last couple of years. I especially thank Sara Simon for providing me with the HFSS simulations used for this analysis. Finally, I thank Peter Hargrave for introducing me to fast telecentric plastic lens refractor designs that can support a wide field of view.

\bibliography{refractor}

\end{document}